\RequirePackage[2024-06-01]{latexrelease}
\documentclass[reprint,amssymb,amsmath,aps,superscriptaddress,floatfix,pra]{revtex4-2}
\RequirePackage{array}
\newcolumntype{C}[1]{>{\centering\arraybackslash}p{#1}} 
\usepackage{epsfig,dsfont,amssymb,amsmath,amsthm,amsfonts,amsbsy,mathrsfs}
\usepackage{graphicx}
\usepackage{booktabs}
\usepackage{multirow}
\usepackage[table,xcdraw]{xcolor}
\usepackage{textcomp}
\usepackage{subfigure}
\usepackage{float}
\usepackage{caption}
\usepackage{booktabs}
\usepackage{epstopdf}
\usepackage[misc]{ifsym}
\usepackage{booktabs}
\usepackage{epstopdf}
\usepackage{upgreek}
\usepackage{dcolumn}
\usepackage{bm}
\usepackage{gensymb}
\usepackage{braket}
\usepackage{sidecap}
\newcommand{\PreserveBackslash}[1]{\let\temp=\\#1\let\\=\temp}

\newcolumntype{C}[1]{>{\PreserveBackslash\centering}p{#1}}
\newcolumntype{R}[1]{>{\PreserveBackslash\raggedleft}p{#1}}
\newcolumntype{L}[1]{>{\PreserveBackslash\raggedright}p{#1}}
\newcolumntype{C}[1]{>{\centering\arraybackslash}p{#1}}
\expandafter\ifx\csname package@font\endcsname\relax\else
\expandafter\expandafter
\expandafter\usepackage
\expandafter\expandafter
\expandafter{\csname package@font\endcsname}
\fi

\newcommand{\be}{\begin{eqnarray}}
\newcommand{\ee}{\end{eqnarray}}
\newcommand{\bfig}{\begin{figure}}
	\newcommand{\efig}{\end{figure}}

\begin{document}
	
	\title{Cryogenic Multiplexing Control Chip for Superconducting Quantum Processor}
	\author{Rutian Huang}
	\email{hrt17@tsinghua.org.cn}
	\affiliation{Laboratory of Superconducting Quantum Information Processing, School of Integrated Circuits, Tsinghua University, Beijing 100084, China}
	\affiliation{Beijing National Research Center for Information Science and Technology, Beijing 100084, China}
	\author{Xiao Geng}
	\affiliation{Laboratory of Superconducting Quantum Information Processing, School of Integrated Circuits, Tsinghua University, Beijing 100084, China}
	\affiliation{Beijing National Research Center for Information Science and Technology, Beijing 100084, China}
	\author{Xinyu Wu}
	\affiliation{Laboratory of Superconducting Quantum Information Processing, School of Integrated Circuits, Tsinghua University, Beijing 100084, China}
	\affiliation{Beijing National Research Center for Information Science and Technology, Beijing 100084, China}
	\author{\\Genting Dai}
	\affiliation{Laboratory of Superconducting Quantum Information Processing, School of Integrated Circuits, Tsinghua University, Beijing 100084, China}
	\affiliation{Beijing National Research Center for Information Science and Technology, Beijing 100084, China}
	\author{Liangliang Yang}
	\affiliation{Laboratory of Superconducting Quantum Information Processing, School of Integrated Circuits, Tsinghua University, Beijing 100084, China}
	\affiliation{Beijing National Research Center for Information Science and Technology, Beijing 100084, China}
	\author{Jianshe Liu}
	\affiliation{Laboratory of Superconducting Quantum Information Processing, School of Integrated Circuits, Tsinghua University, Beijing 100084, China}
	\affiliation{Beijing National Research Center for Information Science and Technology, Beijing 100084, China}
	\author{Wei Chen}
	\email{weichen@tsinghua.edu.cn}
	\affiliation{Laboratory of Superconducting Quantum Information Processing, School of Integrated Circuits, Tsinghua University, Beijing 100084, China}
	\affiliation{Beijing National Research Center for Information Science and Technology, Beijing 100084, China}
	\affiliation{Beijing Innovation Center for Future Chips, Tsinghua University, Beijing 100084, China}
	\date{\today}
	
	
	\begin{abstract}
		This paper proposes a scheme and design of cryogenic multiplexing control chip (MCC) which enables one microwave cable to control many superconducting qubits. Designed based on adjustable inductor, MCC is a non-reciprocal device which generates microwave signals between 4 and 8 GHz and constructs XY control circuit to operate qubits. The effects of filters, isolators, power splitters, qubits and load impedance on MCC are analyzed in the frequency-domain simulation. And in time-domain simulation, the response time, Gaussian pulses, single-sideband, amplitude and frequency compensation of MCC are examined. The MCC can be integrated with the superconducting qubits on the same package at 20 mK temperature stage, and output specific microwave pulses used to drive the superconducting qubits without crosstalks. The MCC can greatly reduce the number of microwave control lines and is beneficial to the development of large-scale superconducting quantum processor.   
	\end{abstract}
	
	\maketitle
	
	
	\section{Introduction}
	The prospect that quantum computing could provide tremendous advantage against classic computer has been emerging with the accomplishments of quantum supremacy by superconducting quantum computers \cite{Arute2019,Gong2021a,Blais2021,Krantz2019,Gong2021a,Harrigan2021,He2021,He2022}. A typical superconducting quantum measurement system usually composed of two parts \cite{Soloviev2021}, one is a superconducting quantum processor \cite{Chen2021}, the other is a classical electronic circuit used to control and readout qubits \cite{AIQuantum2020}. The classical controller at room temperature generates microwave signals \cite{Litvinenko2021}, which pass through meters-long coaxial cables \cite{Krantz2019a},which are expensive and lossy, then go through a series of devices such as attenuators and filters, and finally reaches cryogenic temperature to manipulate qubits \cite{Barends2015,VanMeter2013,Wu2021,AIQuantum2020}. Current systems need a large number of such controling routs \cite{Guarcello2021,Barends2014}. With the development of large scalable quantum computers, however, a key challenge is the interconnect complexity \cite{Litvinenko2021}, due to not only the number of microwave wires will increase, but also the formidable cost \cite{Sliwa2015}, space and cooling power of dilution refrigerator \cite{Huang2021}. A severe interconnect bottleneck \cite{Lim2018,Kerckhoff2015,Kelly2015} therefore appears between the room-temperature control electronics and the quantum processor \cite{Wilen2021}.
	
	To overcome the interconnect bottleneck, an ideal solution is that the electronic controller is designed to operate at the base temperature of refrigerator and integrated with the qubits on the same die or package \cite{Blais2021,Krantz2019,Kandala2021}. Several approaches have been adopted to achieve this goal \cite{Hornibrook2014,Hartmann2021,Abrams2020}. One of the methods is based on placing dissipative semiconductor radio frequency (RF) circuits to multiplex microwave control signals at 10 mK \cite{IEEECircuitsandSystemsSociety,Niffenegger2020}. A number of efforts are underway to use cryogenic complementary metal oxide semiconductor (cryo-CMOS) \cite{Xue2021,Patra2018,Charbon2017,Sebastiano2020} and standard CMOS technology \cite{Arute2019,Prabowo2021,Pauka2021,Sebastiano2020,Xue2021,Charbon2017} to generate control microwave signals located at the 3-4 K stage of dilution refrigerator or lower temperature \cite{Sebastiano2017,Versluis2017,Bardin2019,Patra2018,VanDIjk2020}.
	
	Generally speaking, the electronic controller for superconducting qubits should meet three criteria as follows \cite{Xue2021,Vepsalainen2020,Alexeev2021,Patra2020,Hornibrook2014,Elder2020,Mallet2009,Hashim2020}: 
	(1) Due to the limited cooling power of refrigerator, lower power consumption control systems are necessary.
	(2) Since qubits are very fragile, it is desirable to operate them independently and avoid crosstalk between them.
	(3) For the qubits choosen not to be operated, it can be stored in a stable state without being disturbed by other controlling signals.
	
	In this work, a multiplexing control chip (MCC) is designed for the parallel XY control of superconducting qubits to meet all of the above requirements. MCC is a frequency-domain multiplexing device that works at 20 mK, and it is composed of bandpass filters and tunable isolators, which is non-reciprocal and designed with tunable inductor bridge \cite{Chapman2017}. Here we target qubits that can be resonantly controlled with drive frequencies between 4 and 8 GHz, covering the typical resonance frequencies of superconducting qubits. MCC has functions of attenuation, filter, isolation and reverse isolation and can integrate with the superconducting qubits on the same die or package. It outputs tailored microwave pulses to drive superconducting qubits cooled to 20 mK.
	
	In the following, Section \ref{sec:design&model} presents model analyses and circuit design of MCC. Section \ref{sec:frequency_domain} is the results of frequency-domain simulation, which analyses the effects of filters, isolators, power splitters, qubits and load impedance on MCC. The response time, Gaussian shaped bursts, single-sideband, amplitude and frequency compensation of MCC are examined by the time-domain simulation in Sec. \ref{sec:time_domain}. Conclusions are drawn in Sec. \ref{sec:conclusion}.
	
	
	\section{STRUCTURE DESIGN AND MODEL ANALYSIS}
	\label{sec:design&model}
	
	
	\subsection{Structure Design}
	\label{subsec:Structure Design}
	
\begin{figure}[htb]
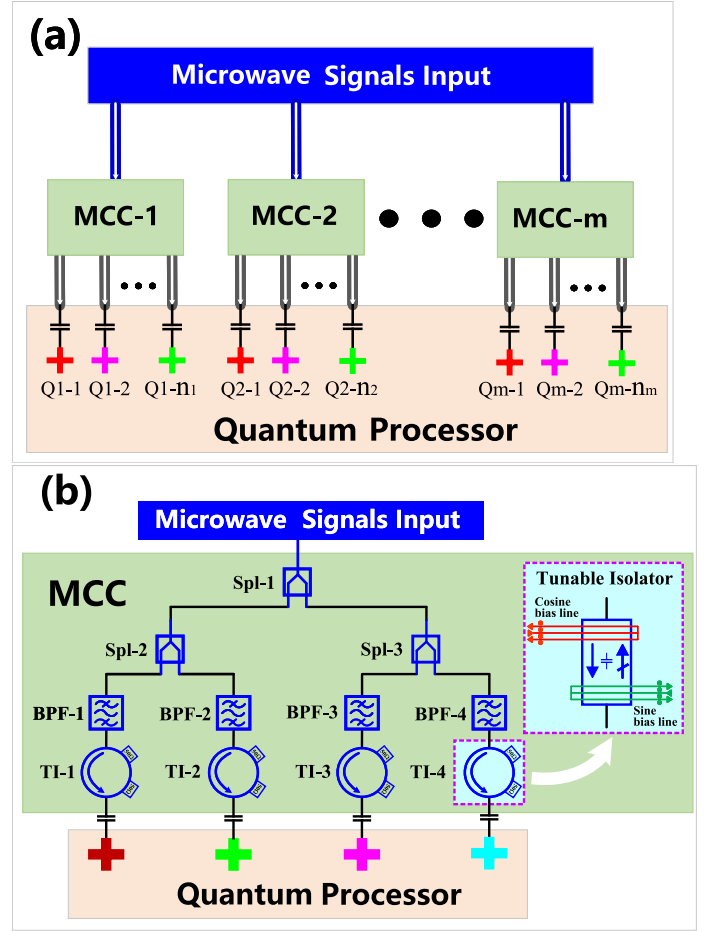

		\begin{center}
			\includegraphics[scale=0.019]{N-channel.pdf}
			\includegraphics[scale=0.018]{four-channel.pdf}
			\caption{Schematic of MCC circuits. (a) Setup of $m$ MCCs in superconducting quantum computer. (b) Circuits diagram of one MCC. The inset is a schematic of a tunable isolator based on SQUIDs circuits.}
			\label{fig:N_four_channel}
		\end{center}
	\end{figure}

	Fig.~\ref{fig:N_four_channel}(a) illustrates several MCCs that are utilized in superconducting quantum computer.  Frequency-domain multiplexed XY control microwave signals are fed into $m$ MCCs, which demultiplex to $\sum_{i=1}^{m}n_i$ microwave controlling signals for the specific qubits on the quantum processor chip. 
	
	For the convenience of circuit analyses, one MCC uses 1:4 proportional control multiplexing, that is, one microwave control line controls four channels of qubits, as shown in Fig.~\ref{fig:N_four_channel}(b). MCC is composed of power splitters, bandpass filters (BPF) and isolators. The power splitters divide microwaves into different channels. The BPFs have the same bandwidth but different central frequencies to obtain signals of specific frequencies for qubits. Table I lists bandwiths and center frequencies of BPFs in MCC. The isolators are used to select the qubit to be manipulated and avoid crosstalk from other qubits. Designed with tunable inductor bridge, the non-reciprocity of the isolator is realized by connecting matching loads to two of the four-port on-chip superconducting circulator \cite{Chapman2017,Kerckhoff2015}. The center frequency of the isolator can be precisely set to allow signal for only one qubit to pass by adjusting the capacitance in the isolator when it is designed or by changing current in its bias lines when it is at work \cite{Chapman2017,Kerckhoff2015}. As shown in the inset of Fig.~\ref{fig:N_four_channel}(b), two bias lines (cosine and sine) are used to adjust the magnetic field on which the isolator is located.  According to Auster's principle, the center frequency of the isolator can be changed by adjusting the current passing through the bias line and changing the magnetic flux passing through the  superconducting quantum interference devices (SQUIDs), thus altering the equivalent inductance of the SQUIDs. In this way, the signals for each qubit are transmitted independently without interference.    
	
	The room-temperature control circuits send signals via MCC to manipulate qubits. MCC can accurately divide multiple frequency microwave signals into specific channels so as to independently control and operate qubits to avoid crosstalk among different channels. To protect the quantum information of other channels from interference, the quantum signal from qubits should not reflect back through MCC to other routs, thanks to the non-reciprocity of the tunable isolators in MCC. By reversing the direction of some isolators, the MCC can be free to choose not to operate specific qubits, which can store in a steady state without being disturbed. 
	
	\begin{figure}[htb]
		\begin{center}
			\includegraphics[scale=0.017]{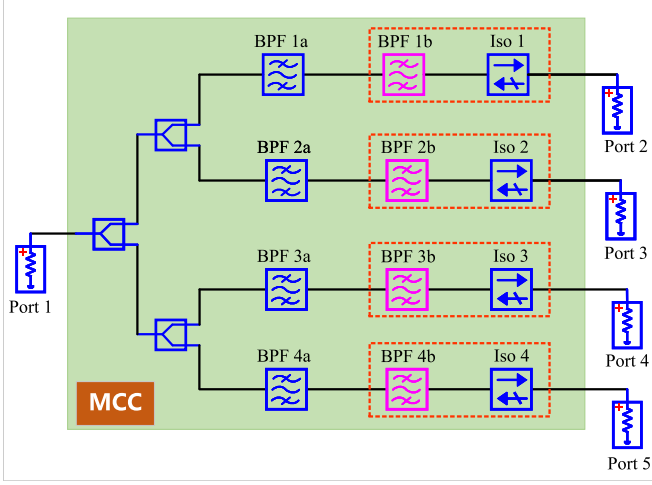}
			\caption {Circuit diagram of MCC for frequency-domain simulation.}
			\label{fig:frequency-simulation}
		\end{center}
	\end{figure}
	
	
	\subsection{Frequency-domain model analysis}
	\label{subsec:Frequency-domain model analysis}
	
	Fig.~\ref{fig:frequency-simulation} displays the circuit diagram of MCC for frequency-domain simulation with Advanced Design System© software. In order to realize the non-reciprocity and accurate filtering function of on-chip isolator, the combination of filter and isolator is adopted, which is shown in the red dashed box in Fig.~\ref{fig:frequency-simulation}. In this circuit model, BPF 1a-4a are the designed filters which have the same bandwidth but different center frequencies corresponding to the bandpass filters BPF 1-4 in Fig.~\ref{fig:N_four_channel}(b) respectively. The intrinsic finite bandwith of the actual tunable isolators are captured by bandpass filters BPF 1b-4b, which is consistent with Bessel feature \cite{Chapman2017,Kerckhoff2015}, and Iso 1-4 are the ideal isolators with infinite bandpass in Fig.~\ref{fig:frequency-simulation}. Therefore, the tunable isolators TI-1-4 in Fig.~\ref{fig:N_four_channel}(b) are the combination of BPF 1b-4b and Iso 1-4, respectively.
	
	The scatter parameter matrix of MCC among the five ports is as follows:
	\begin{equation}
	\left[
	\begin{array}{c}
	b_{1,\rm {out}}(t)\\
	b_{2,\rm {out}}(t)\\
	b_{3,\rm {out}}(t)\\
	b_{4,\rm {out}}(t)\\
	b_{5,\rm {out}}(t)
	\end{array}
	\right]=
	\left[
	\begin{array}{ccccc}
	\mathcal{S}_{11} & \mathcal{S}_{12} & \mathcal{S}_{13} & \mathcal{S}_{14} & \mathcal{S}_{15}\\
	\mathcal{S}_{21} & \mathcal{S}_{22} & \mathcal{S}_{23} & \mathcal{S}_{24} & \mathcal{S}_{25}\\
	\mathcal{S}_{31} & \mathcal{S}_{32} & \mathcal{S}_{33} & \mathcal{S}_{34} & \mathcal{S}_{35}\\
	\mathcal{S}_{41} & \mathcal{S}_{42} & \mathcal{S}_{43} & \mathcal{S}_{44} & \mathcal{S}_{45}\\
	\mathcal{S}_{51} & \mathcal{S}_{52} & \mathcal{S}_{53} & \mathcal{S}_{54} & \mathcal{S}_{55}
	\end{array}
	\right]\left[
	\begin{array}{c}
	a_{1,\rm {in}}(t)\\
	a_{2,\rm {in}}(t)\\
	a_{3,\rm {in}}(t)\\
	a_{4,\rm {in}}(t)\\
	a_{5,\rm {in}}(t)
	\end{array}
	\right],
	\end{equation}
	
	\begin{equation}
	S_{ij}= 10\lg {\mathcal{S}_{ij}},
	\end{equation}
	where $S_{ij}$ and $\mathcal{S}_{ij}$ ($i$=1-5, $j$=1-5) both indicate the transmission characteristic parameter from port $j$ to port $i$, $a_{i,\rm in}(t)$ and $b_{j,\rm out}(t)$ represent the input and output ports parameters, respectively. $S_{ij}$ is in dB unit, while $\mathcal{S}_{ij}$ is in linear unit.
	
	The ideal scatter parameter matrix is given by
	\begin{equation}
	\label{ideal-s}
	\left[
	\begin{array}{c}
	b_{1,\rm {out}}(t)\\
	b_{2,\rm {out}}(t)\\
	b_{3,\rm {out}}(t)\\
	b_{4,\rm {out}}(t)\\
	b_{5,\rm {out}}(t)
	\end{array}
	\right]=
	\left[
	\begin{array}{ccccc}
	\mathcal{S}_{11} & 0 & 0 & 0 & 0\\
	\mathcal{S}_{21} & 0 & 0 & 0 & 0\\
	\mathcal{S}_{31} & 0 & 0 & 0 & 0\\
	\mathcal{S}_{41} & 0 & 0 & 0 & 0\\
	\mathcal{S}_{51} & 0 & 0 & 0 & 0
	\end{array}
	\right]\left[
	\begin{array}{c}
	a_{1,\rm {in}}(t)\\
	a_{2,\rm {in}}(t)\\
	a_{3,\rm {in}}(t)\\
	a_{4,\rm {in}}(t)\\
	a_{5,\rm {in}}(t)
	\end{array}
	\right],
	\end{equation}
	where $\mathcal{S}_{i1}$ ($i$=1-5) represents the transmission feature parameters from port 1 to other ports. In the ideal scatter parameter matrix, all parameters except $\mathcal{S}_{i1}$ are 0, i.e., there is no crosstalk between different channels. In frequency domain analysis, transmission characteristic parameters and crosstalk between different ports are obtained by simulating $\mathcal{S}_{ij}$ parameters.    
	
	
	\subsection{Time-domain model analysis}
	\label{subsec:Time-domain model analysis}
	
	\begin{figure}[htb]
		\begin{center}
			\includegraphics[scale=0.02]{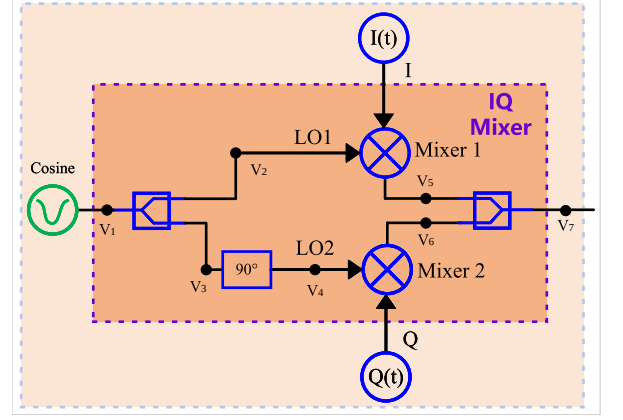}
			\caption {Circuit diagram of IQ mixer in MCC in time domain simulation.}
			\label{fig:IQ-Mixer}
		\end{center}
	\end{figure}
	
	In time-domain simulation of MCC, single-sideband (SSB) microwave signals are generated by using IQ mixers. The IQ mixer can be represented by two mixers and two hybrid couplers \cite{Raab2015} or power splitters. Fig.~\ref{fig:IQ-Mixer} depicts the working principle of IQ mixer. In IQ mixer, the function formula used for $I(t)$ and $Q(t)$ are
	\begin{equation}
	I(t)= \cos(\omega_{\rm{IF}}t),
	\end{equation}
	\begin{equation}
	Q(t)= \sin(\omega_{\rm{IF}}t),
	\end{equation}
	where the intermediate frequency $\omega_{\rm{IF}}= 2 \pi f_i$, $i=$1-4, 
	$f_1=100$ \rm{MHz},
	$f_2=300$ \rm{MHz},
	$f_3=500$ \rm{MHz},
	$f_4=700$ \rm{MHz},
	$t=0-850$ \rm{ns} with step size 10 ns.
	
	\begin{figure*}[htb]
		\begin{center}
			\includegraphics[scale=0.033]{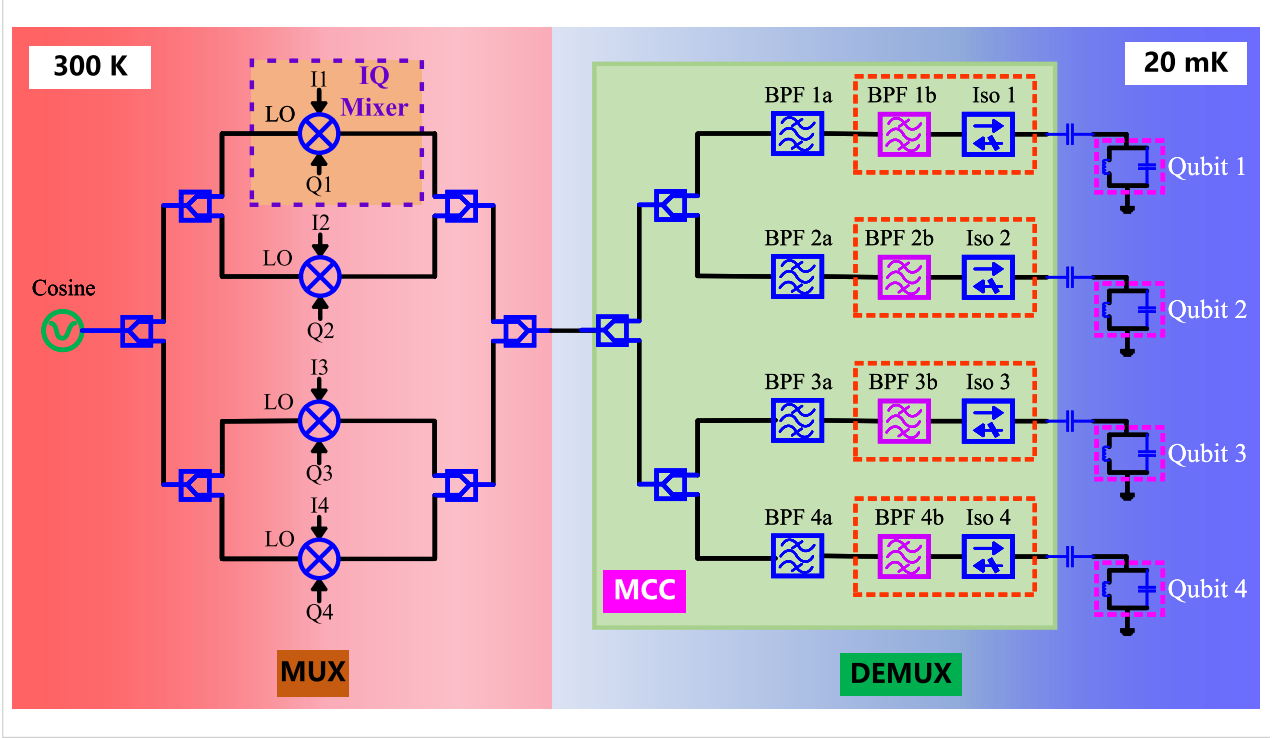}
			\caption {Circuit diagram of MCC in time domain simulation.}
			\label{fig:domain-simulation}
		\end{center}
	\end{figure*}
	
	As shown in Fig.~\ref{fig:IQ-Mixer}, when the phase shift is $90\degree$, upper sideband (USB) signal is generated as follows:
	\begin{equation}
	V_1(t)=\cos(\omega_{\rm{LO}}t),
	\end{equation}
	\begin{equation}
	V_2(t)=V_3(t)=\frac{M}{2}\cos(\omega_{\rm{LO}}t),
	\end{equation}
	\begin{equation}
	V_4(t)=\frac{M}{2}\cos(\omega_{\rm{LO}}t+\frac{\pi}{2}),
	\end{equation}
	\begin{equation}
	\begin{aligned}
	V_5(t)&=KV_2(t)I(t)\\
	&=K\frac{M}{2}\cos(\omega_{\rm{LO}}t)\cos(\omega_{\rm{IF}}t),
	\end{aligned}
	\end{equation}
	\begin{equation}
	\begin{aligned}
	V_6(t)&=KV_4(t)Q(t)\\
	&=K\frac{M}{2}\cos(\omega_{\rm{LO}}t+\frac{\pi}{2})\sin(\omega_{\rm{IF}}t),
	\end{aligned}
	\end{equation}
	\begin{equation}
	\begin{aligned}
	V_7(t)&=V_5(t)+V_6(t)\\
	&=K\frac{M}{2}\cos(\omega_{\rm{LO}}t)\cos(\omega_{\rm{IF}}t)\\
	&+K\frac{M}{2}\cos(\omega_{\rm{LO}}t+\frac{\pi}{2})\sin(\omega_{\rm{IF}}t)\\
	&=\frac{KM}{2}\cos(\omega_{\rm{LO}}+\omega_{\rm{IF}})t,
	\end{aligned}
	\end{equation}
	where $\omega_{\rm{LO}}$ is the frequency of local oscillator (LO), $V_i(t)$ ($i$=1-7) are the voltages on a series of nodes, $M$ is the attenuation coefficient of power splitters, and $K$ is a constant accounting for the voltage conversion loss of the mixer \cite{Pozar2005,Raab2015}.
	
	When the phase shift is $-90\degree$, lower sideband (LSB) signal is generated as follows:
	\begin{equation}
	V_4(t)=\frac{M}{2}\cos(\omega_{\rm{LO}}t-\frac{\pi}{2}),
	\end{equation}
	\begin{equation}
	\begin{aligned}
	V_5(t)&=KV_2(t)I(t)\\
	&=K\frac{M}{2}\cos(\omega_{\rm{LO}}t)\cos(\omega_{\rm{IF}}t),
	\end{aligned}
	\end{equation}
	\begin{equation}
	\begin{aligned}
	V_6(t)&=KV_4(t)Q(t)\\
	&= K\frac{M}{2}\cos(\omega_{\rm{LO}}t-\frac{\pi}{2})\sin(\omega_{\rm{IF}}t),
	\end{aligned}
	\end{equation}
	\begin{equation}
	\begin{aligned}
	V_7(t)&=V_5(t)+V_6(t)\\
	&=K\frac{M}{2}\cos(\omega_{\rm{LO}}t)\cos(\omega_{\rm{IF}}t)\\
	&+K\frac{M}{2}\cos(\omega_{\rm{LO}}t-\frac{\pi}{2})\sin(\omega_{\rm{IF}}t)\\
	&=\frac{KM}{2}\cos(\omega_{\rm{LO}}-\omega_{\rm{IF}})t.
	\end{aligned}
	\end{equation}
	
	Thus, the IQ mixer is employed to generate the desired USB or LSB signals. 
	
	Fig.~\ref{fig:domain-simulation} shows the system-level circuit diagram of MCC in the time-domain simulation, where SSB microwave signals are employed. In the left red box, a microwave source of single-tone is send to the local oscillator (LO) signal ports of four IQ Mixers via power splitters. By adjusting the $I_i(t)$ and $Q_i(t)$ ($i$=1-4) input signals of IQ mixers, four SSB microwave signals are produced. The synthesizing signals are combined into one microwave line through power splitters, thus completing the multiplexing (MUX) process. In the right box is the MCC device, which demultiplexes (DEMUX) four signal-tones into four channels for independent control of four qubits.
	
	
	\section{FREQUENCY-DOMAIN SIMULATION}
	\label{sec:frequency_domain}
	
	\begin{figure*}[htb]
		\begin{center}
			\includegraphics[scale=0.35]{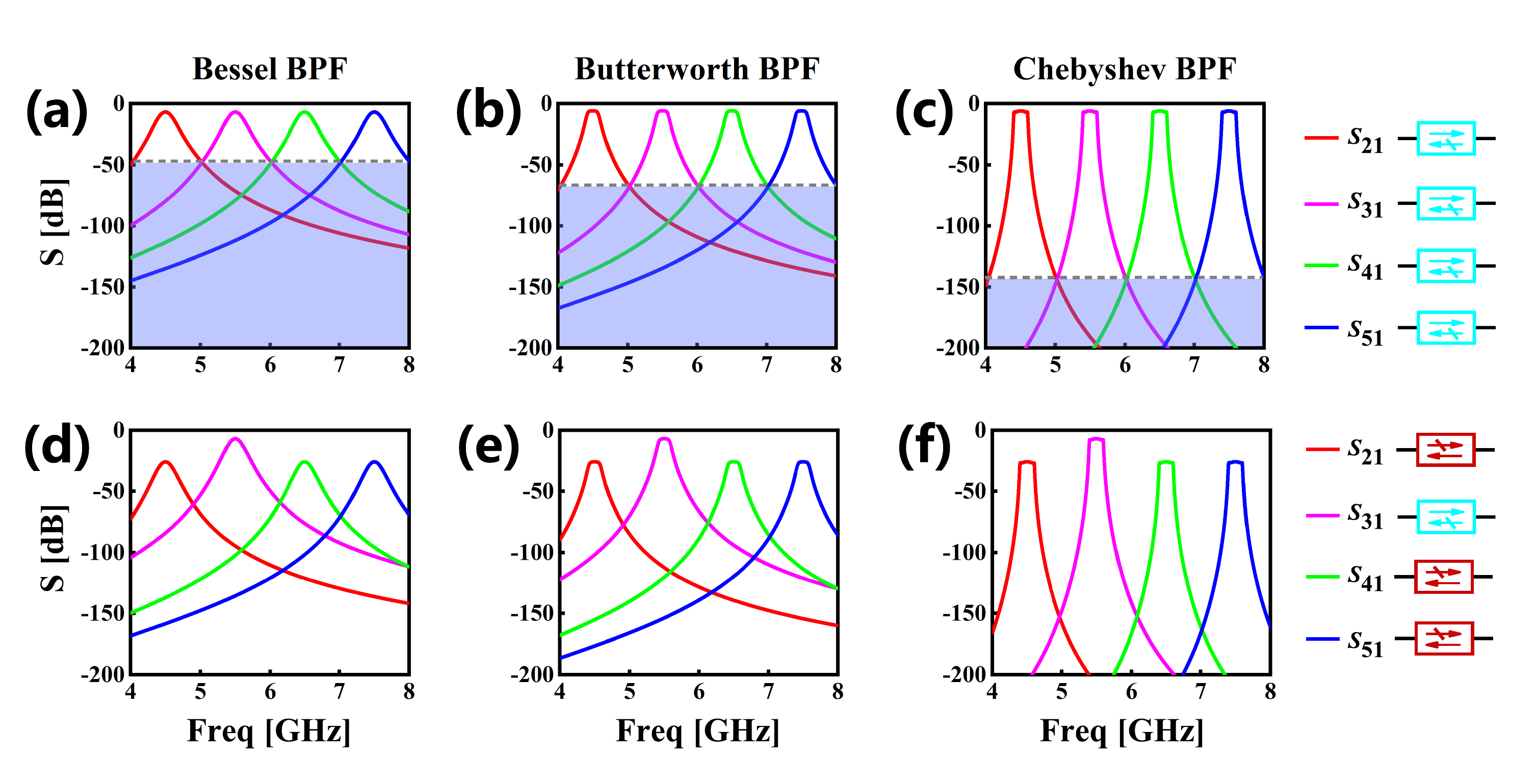}
			\caption {Comparison of S parameters of MCC with three different kinds of BPF. The isolator symbols on the right of (a)(b)(c) indicate that all the four channels are accessible, (a) Bessel BPF, (b) Butterworth BPF, (c) Chebyshev BPF, where the gray dashed lines mark the overlapping frequencies. The isolator symbols on the right of (d)(e)(f) indicate that only the second rout (center frequency 5.5 GHz) is accessible.}
			\label{fig:F-filters}
		\end{center}
	\end{figure*}
	
	In the frequency-domain simulation of MCC, the effects of filters, isolators, power splitters, load impedance and qubits, and crosstalk on the scattering parameters of MCC are examined, respectively.
	
	
	\subsection{Effects of Filters}
	\label{subsec:F_filters}
	
	To investigate the effects of filters on the performance of MCC, the parameters of isolators are fixed with an isolation degree of 20 dB, standing wave ratio of 1.2, and insertion loss of 0.9 dB.
	
	As mentioned in Sec. \ref{subsec:Frequency-domain model analysis} and shown in Fig.~\ref{fig:frequency-simulation}, there are two sets of fliters in the MCC, i.e., BPF 1a-4a as one set, BPF 1b-4b as the other. Table \ref{tab:BPF} lists the center frequencies ($f_0$) and bandwiths of BPFs, which include three types of BPFs, i.e., Bessel BPF, Butterworth BPF and Chebyshev BPF. The bandwidths are 0.5 GHz when Filter 1a-4a are set as Bessel filters, and 0.2 GHz of BPF 1b-4b, which are Bessel BPFs. Whlie the bandwidths are 0.2 GHz when Filter 1a-4a are set as Butterworth filters or Chebyshev filters, and 0.3 GHz of BPF 1b-4b, which are also Bessel BPFs.
	
	\begin{table*}[]
		\renewcommand\arraystretch{2}
		\centering
		\caption{Bandwidths and center frequencies ($f_0$) of BPFs in MCC.}
		\label{tab:BPF}
		\begin{tabular}{C{1.6cm}C{2cm}C{2cm}C{2cm}C{2cm}C{2cm}C{2cm}}
			\hline
			\hline
			& \multicolumn{2}{c}{Bessel}  & \multicolumn{2}{c}{Butterworth} & \multicolumn{2}{c}{Chebyshev} \\
			\begin{tabular}[c]{@{}c@{}}BPF\end{tabular} &
			\begin{tabular}[c]{@{}c@{}}$f_0$(GHz)\end{tabular} &
			\begin{tabular}[c]{@{}c@{}}Bandwith(GHz)\end{tabular} &
			\begin{tabular}[c]{@{}c@{}}$f_0$(GHz)\end{tabular} &
			\begin{tabular}[c]{@{}c@{}}Bandwith(GHz)\end{tabular} &
			\begin{tabular}[c]{@{}c@{}}$f_0$(GHz)\end{tabular} &
			\begin{tabular}[c]{@{}c@{}}Bandwith(GHz)\end{tabular} \\ 
			\hline
			1a & 4.5          & 0.5          & 4.5                 & 0.2       & 4.5                & 0.2      \\
			2a & 5.5          & 0.5          & 5.5                 & 0.2       & 5.5                & 0.2      \\
			3a & 6.5          & 0.5          & 6.5                 & 0.2       & 6.5                & 0.2      \\
			4a & 7.5          & 0.5          & 7.5                 & 0.2       & 7.5                & 0.2      \\  
			1b & 4.5          & 0.2          & 4.5                 & 0.3       & 4.5                & 0.3      \\
			2b & 5.5          & 0.2          & 5.5                 & 0.3       & 5.5                & 0.3      \\
			3b & 6.5          & 0.2          & 6.5                 & 0.3       & 6.5                & 0.3      \\
			4b & 7.5          & 0.2          & 7.5                 & 0.3       & 7.5                & 0.3      \\ \hline
			\hline
		\end{tabular}
	\end{table*}
	
	\begin{figure*}[htb]
		\begin{center}
			\includegraphics[scale=0.21]{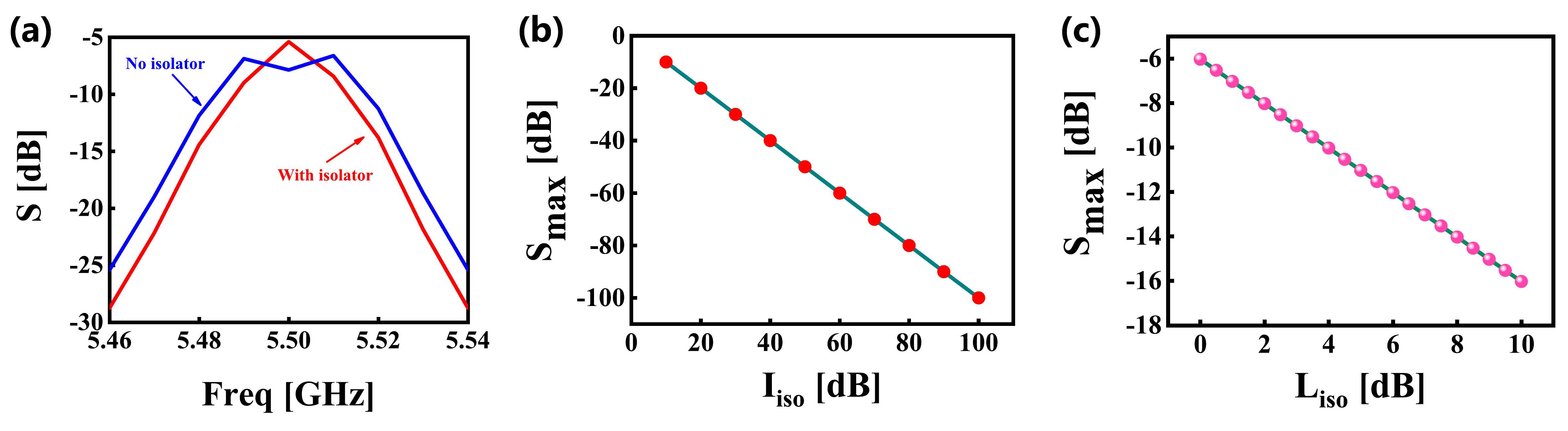}
			\caption {The effects of isolators on MCC. (a) When the impedances are not matched, there is only one peak (solid red line) if the isolators are used in MCC, and there are two peaks (solid blue line) on the curve without isolators. (b) The relationship between $S_{\rm{max}}$ and the isolation degree $\mathcal{I}_{\rm{iso}}$ of Iso 2. (c) The variation of $S_{\rm{max}}$ with the insertion loss $\mathcal{L}_{\rm{iso}}$ of Iso 2.}
			\label{fig:F-isolator}
		\end{center}
	\end{figure*}
	
	In order to choose appropriate filters as Filter 1a-4a, the S parameters of three kinds of circuits with Filter 1a-4a set as Bessel BPF, Butterworth BPF and Chebyshev BPF are simulated. Fig.~\ref{fig:F-filters} gives comparison of S parameters of MCC with three different types of BPF 1a-4a. The isolator symbols on the right of Fig.~\ref{fig:F-filters} (a)(b)(c) indicate that all the four channels access. It can be seen clearly that MCC can realize the function of control multiplexing by using these three BPFs, and the signals of different channels do not interfere with each other. As marked by the dashed lines in Fig.~\ref{fig:F-filters} (a)(b)(c), the S parameters at the overlapped frequencies are $-50$, $-67$, $-142$ dB, respectively. The flat tops of the S parameters in Fig.~\ref{fig:F-filters} (c) suggest that Chebyshev BPF can provide the robustness of the control signals within the bandwith. At the frequencies out of the passband, the curves of S parameters in Fig.~\ref{fig:F-filters} (c) are much steeper than that of Fig.~\ref{fig:F-filters}(a) and Fig.~\ref{fig:F-filters} (b), which means MCCs with Chebyshev filters as Filter 1a-4a can prevent interference from signal of frequencies out of the corresponding qubit (Qubit 1-4) frequencies better than two other kinds of MCC (Bessel and Butterworth) do. These indicate Chebyshev BPFs have the best performance of frequency selection.  
	
	Besides being able to multiplexing one microwave channel, MCC can also decide whether to block control signal for a specific qubit by turning over the direction of the corresponding isolator. In reality, turning over the direction of isolator can be achieved by reversing the direction of the magnetic field (provided by bias lines) through the SQUIDs in the on-chip isolator \cite{Chapman2017,Kerckhoff2015}. To show the effect of isolation, a simulation is carried out where all qubits are shielded from control signal except Qubit 2.
	
	In Fig.~\ref{fig:F-filters} (d)(e)(f), the isolator symbols on the right indicate that only the second rout (center frequency 5.5 GHz) is accessible.  The S parameters of MCC with different filters in this case are shown. It can be seen that the maximum of $S_{21}$, $S_{41}$ and $S_{51}$ (corresponding to Qubit 1, 3 and 4) are below the value of $S_{31}$ at the frequency of Qubit 2, which is close to 0 dB. This means only Qubit 2 can be contolled but other qubits are independent from control because of isolation of respective control signal. The maximum of $S_{21}$, $S_{41}$ and $S_{51}$ in Fig.~\ref{fig:F-filters} (d) are over $-50$ dB, much higher than those of Fig.~\ref{fig:F-filters} (e) and (f). Therefore, MCC with Chebyshev or Butterworth filters isolate qubits better than MCC with Bessel filters.
	
	Considering the results of comparison above, Chebyshev BPF is the most appropriate filter for Filter 1a-4a, and is therefore chosen for the following simulations of MCC.
	
	
	\subsection{Effects of Isolators} 
	\label{subsec:F_isolators}
	
	To justify the necessity of isolators in the MCC, its S parameters are simulated with or without isolators, respectively. 
	
	When impedances are matched, the S parameters of MCC are the same in the cases with or without isolators. While the impedances are not matched, as shown in  Fig.~\ref{fig:F-isolator} (a), there is only one peak if the isolators are used, meaning that the signal can be precisely routed into the channel (solid red line). In the case without isolators, there are two peaks indicating that the MCC cannot accurately rout into the specific channel (solid blue line). Therefore, isolators are necessary for the design of MCC. 
	
	Isolation degree and insertion loss of isolator are the two most important indexes. Next, the influence of isolation degree and insertion loss on MCC is discussed. 
	
	MCC is set so that only the second channel in Fig.~\ref{fig:frequency-simulation} is chosen, while the other channels are blocked by reversing the direction of isolators. Fig.~\ref{fig:F-isolator} (b) gives the relationship between $S_{\rm{max}}$, which is the maximum of ${S}_{21}$, and the isolation degree $\mathcal{I}_{\rm{iso}}$ of Iso 2. The curve can be fitted by a linear function
	\begin{equation}
	S_{\rm{max}}=-\mathcal{I}_{\rm{iso}}.  
	\end{equation}
	
	Fig.~\ref{fig:F-isolator} (c) plots the variation of $S_{\rm{max}}$ with the insertion loss $\mathcal{L}_{\rm{iso}}$ of Iso 2. Fitting the curve yields
	\begin{equation}
	S_{\rm{max}}=-\mathcal{L}_{\rm{iso}}-6.023,
	\end{equation}
	where $-6.023$ dB is the loss caused by the power splitter.
	
	
	\subsection{Effects of power splitter}
	\label{subsec:F_splitter}
	
	\begin{table*}[]
		\renewcommand\arraystretch{2}
		\centering
		\caption{The effects of power splitters on MCC.}
		\label{tab:power splitter}
		\begin{tabular}{cccllll}
			\hline
			\hline
			\multirow{3}{*}{$\rm{S_{max}}$} & \multicolumn{2}{c}{Bessel BPF}      & \multicolumn{2}{c}{Butterworth BPF}  & \multicolumn{2}{c}{Chebyshev BPF}    \\ \cline{2-7} 
			&
			\begin{tabular}[c]{@{}c@{}}no splitter\end{tabular} &
			\begin{tabular}[c]{@{}c@{}}with splitter\end{tabular} &
			\multicolumn{1}{c}{\begin{tabular}[c]{@{}c@{}}no splitter\end{tabular}} &
			\multicolumn{1}{c}{\begin{tabular}[c]{@{}c@{}}with splitter\end{tabular}} &
			\multicolumn{1}{c}{\begin{tabular}[c]{@{}c@{}}no splitter\end{tabular}} &
			\multicolumn{1}{c}{\begin{tabular}[c]{@{}c@{}}with splitter\end{tabular}} \\ 
			& \multicolumn{1}{c}{\begin{tabular}[c]{@{}c@{}}(GHz, dB)\end{tabular}} 
			& \multicolumn{1}{c}{\begin{tabular}[c]{@{}c@{}}(GHz, dB)\end{tabular}}
			& \multicolumn{1}{c}{\begin{tabular}[c]{@{}c@{}}(GHz, dB)\end{tabular}}
			& \multicolumn{1}{c}{\begin{tabular}[c]{@{}c@{}}(GHz, dB)\end{tabular}}
			& \multicolumn{1}{c}{\begin{tabular}[c]{@{}c@{}}(GHz, dB)\end{tabular}}
			& \multicolumn{1}{c}{\begin{tabular}[c]{@{}c@{}}(GHz, dB)\end{tabular}}\\
			\hline
			$\rm{m_1}$                    & (4.504,$-0.919$)  & (4.497,$-6.924$) & (4.502,$-0.002$)  & (4.498,$-6.023$) & (4.508,$-0.007$)  & (4.497,$-6.023$) \\
			$\rm{m_2}$                    & (5.499,$-0.879$) & (5.498,$-6.923$) & (5.499,$-0.0001$) & (5.499,$-6.023$) & (5.500,$-0.007$)  & (5.498,$-6.023$) \\
			$\rm{m_3}$                    & (6.495,$-0.884$) & (6.498,$-6.923$) & (6.497,$-0.0001$) & (6.499,$-6.023$) & (6.492,$-0.0008$) & (6.498,$-6.023$) \\
			$\rm{m_4}$                    & (7.490,$-0.931$) & (7.498,$-6.923$) & (7.495,$-0.006$)  & (7.499,$-6.023$) & (7.488,$-0.020$)  & (7.499,$-6.023$) \\ 
			\hline
			\hline
		\end{tabular}
	\end{table*}
	
	\begin{figure}[htb]
		\begin{center}
			\includegraphics[scale=0.23]{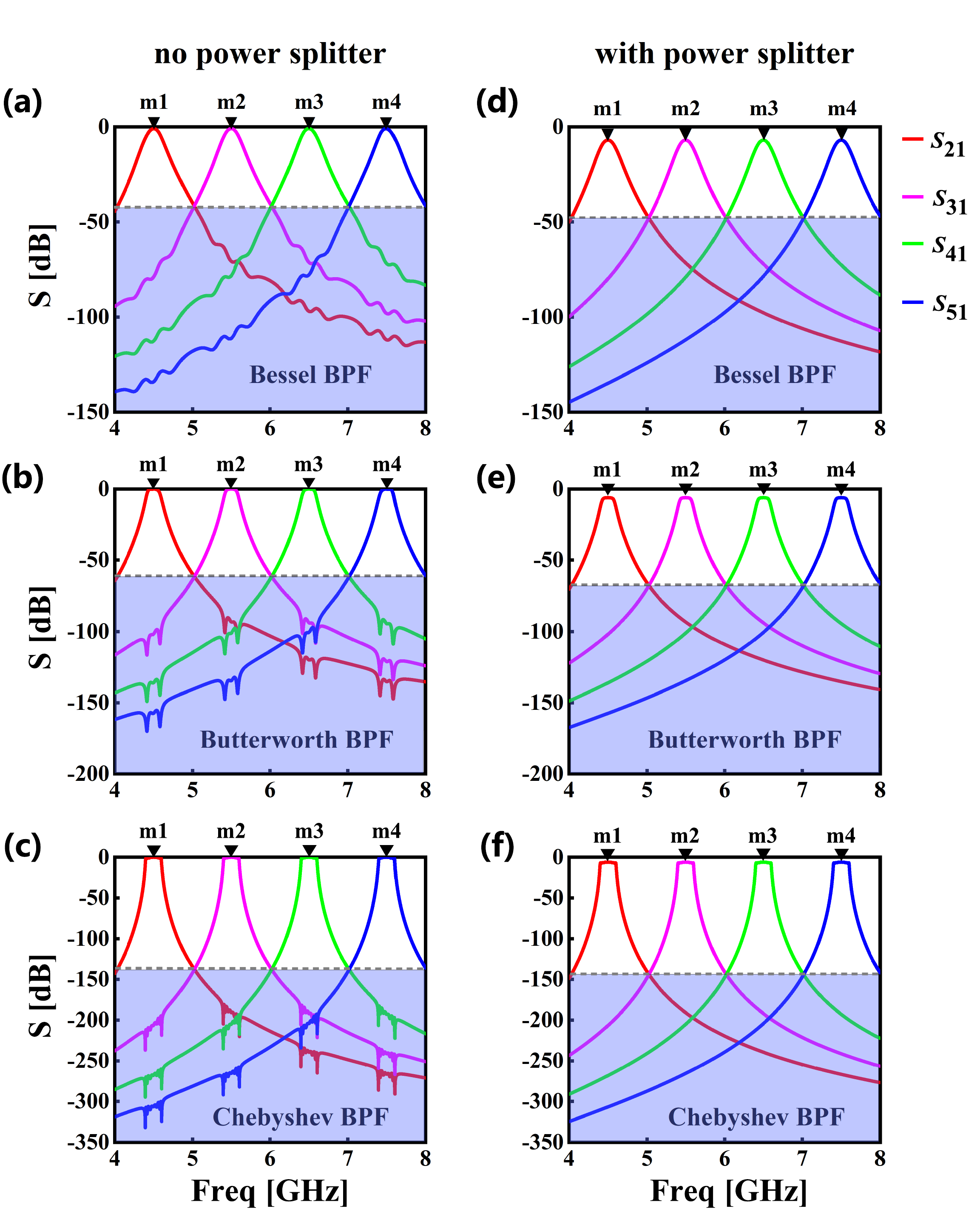}
			\caption{The influence of power splitters on MCC. (a) With power splitters and Bessel BPF, (b) with power splitters and Butterworth BPF, (c) with power splitters and Chebyshev BPF, (d) without power splitters and Bessel BPF, (e) without power splitters and Butterworth BPF, (f) without power splitters and Chebyshev BPF.}
			\label{fig:F-power-splitter}
		\end{center}
	\end{figure}
	
	The S parameters of MCC are simulated for the circuit with or without power splitters.
	Fig.~\ref{fig:F-power-splitter} (a)(b)(c) present the simulation results with power splitters for the three different filters, i.e., Bessel BPF, Butterworth BPF and Chebyshev BPF, respectively. In comparison, Fig.~\ref{fig:F-power-splitter} (d)(e)(f) give the simulation results without power splitters. It is noticeable that for each kind of BPFs, S parameters on the right panel (with power splitter) are lower than that of the left panal (without power splitter). The maximums ($m_1$, $m_2$, $m_3$, $m_4$) of the S parameters in Fig.~\ref{fig:F-power-splitter} are listed in Tab \ref{tab:power splitter}.
	
	As marked by the dashed lines in Fig.~\ref{fig:F-power-splitter} (a)(b)(c), the S parameters at the overlapped frequencies are $-50$, $-67$, $-142$ dB, respectively, and in Fig.~\ref{fig:F-power-splitter} (d)(e)(f) the S parameters at the overlapped frequencies are $-56$, $-73$, $-148$ dB. Thus, the S paramaters at the overlapped frequencies for the circuits with power splitters are 6 dB lower than that without power splitters. Because attenuators are needed in the superconducting quantum test system, the 6 dB attenuation of the power splitters should be taking into account.
	
	The curves of S parameters above the dashed lines in Fig.~\ref{fig:F-power-splitter} (d)(e)(f) are nearly the same as that of Fig.~\ref{fig:F-power-splitter} (a)(b)(c). Beneath the dashed lines, however, the curves of S parameters in Fig.~\ref{fig:F-power-splitter} (d)(e)(f) are much smoother than that of Fig.~\ref{fig:F-power-splitter} (a)(b)(c). Consequently, the power splitters are used in MCC in the following simulations.
	
	\begin{figure}[htb]
		\begin{center}
			\includegraphics[scale=0.3]{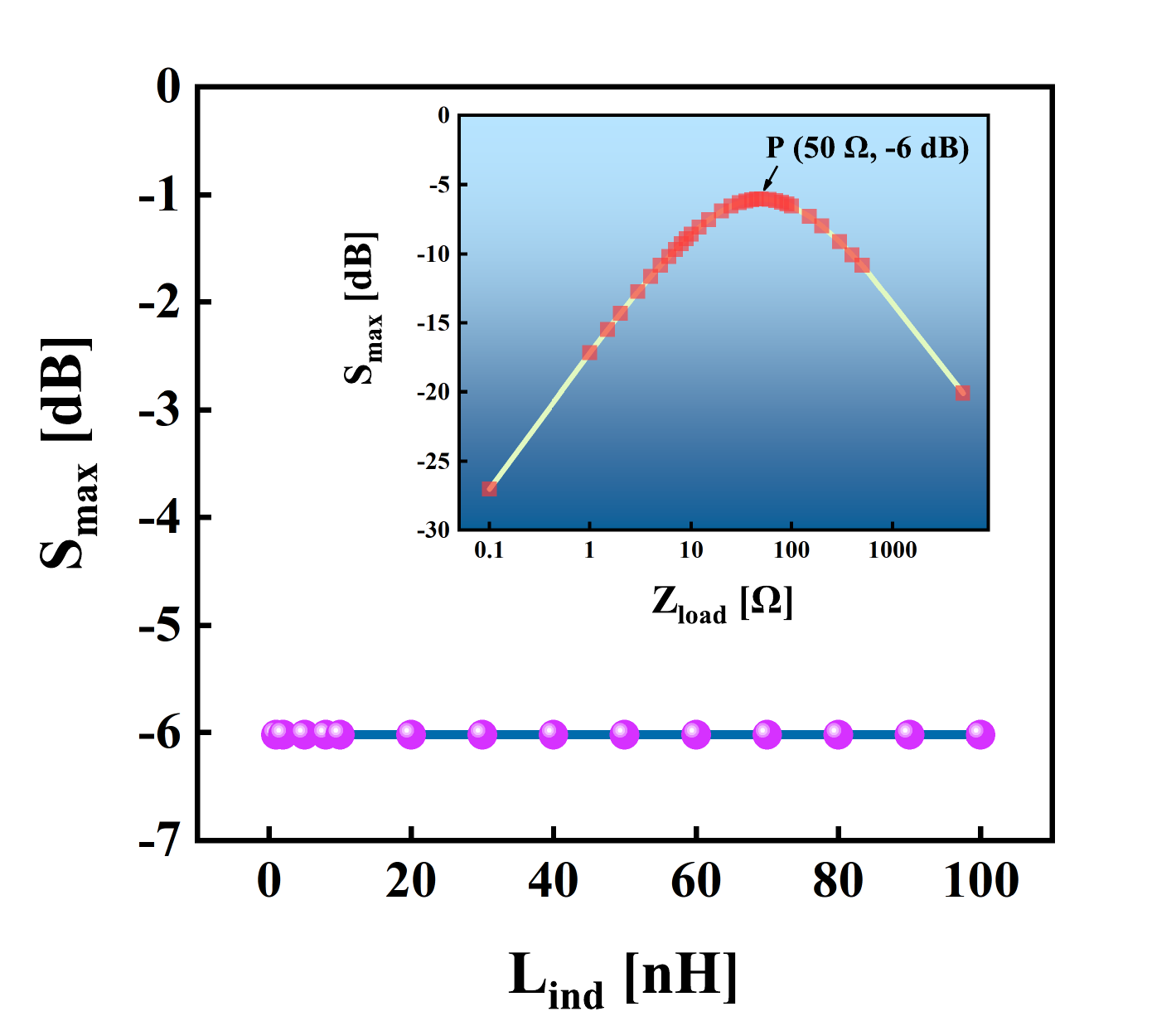}
			\caption{The relationship between $S_{\rm{max}}$ and the inductance of qubits at the matching load impedance of 50 $\Omega$. The inset is a plot of $S_{\rm{max}}$ vs load impedance $Z_{\rm{load}}$.}
			\label{fig:F-qubits-impedance}
		\end{center}
	\end{figure}
	
	\begin{figure*}[htbp]
		\begin{center}
			\includegraphics[scale=0.25]{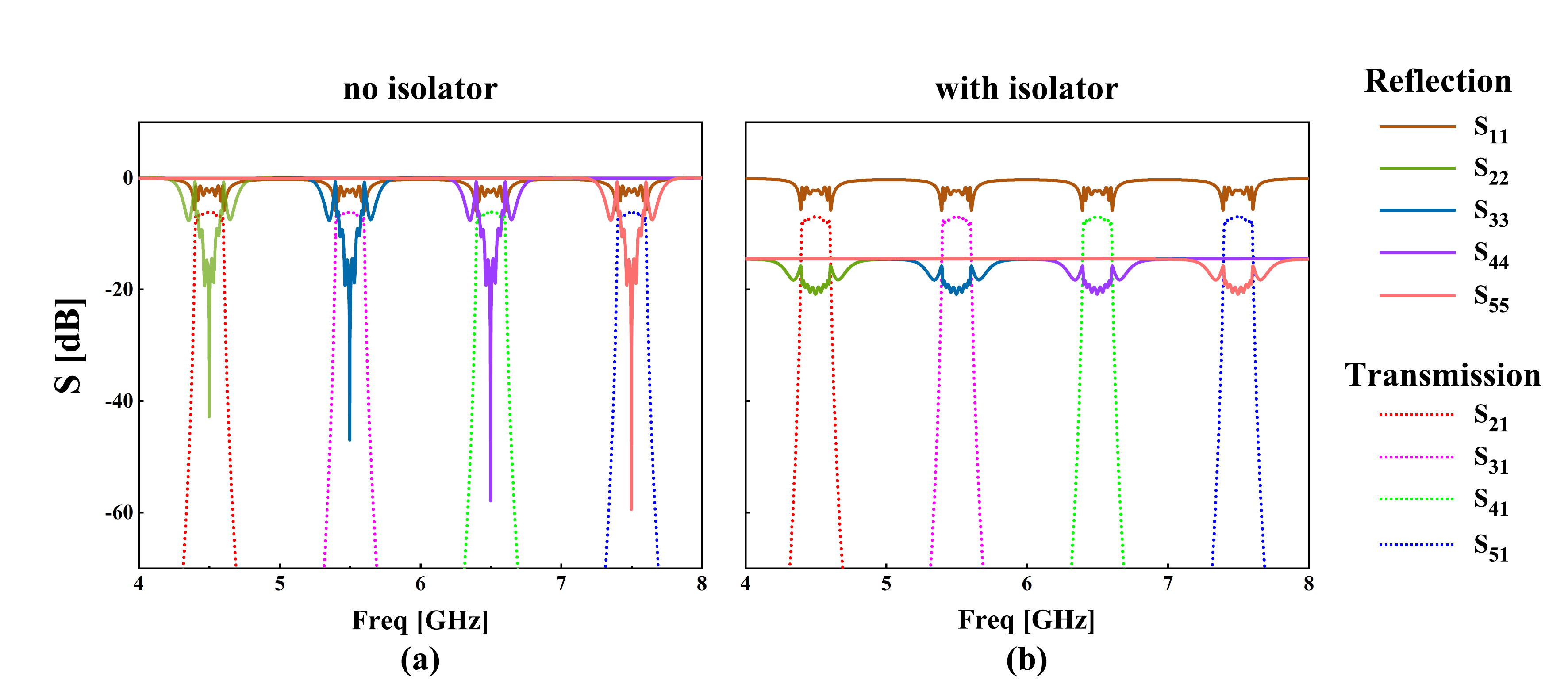}
			\caption{Reflections of MCC without isolators (a) and with isolators (b).}
			\label{fig:F-reflection}
		\end{center}
	\end{figure*}
	
	\begin{figure*}[htbp]
		\centering
		\subfigure
		{\includegraphics[height=4.6cm,width=7.2cm]{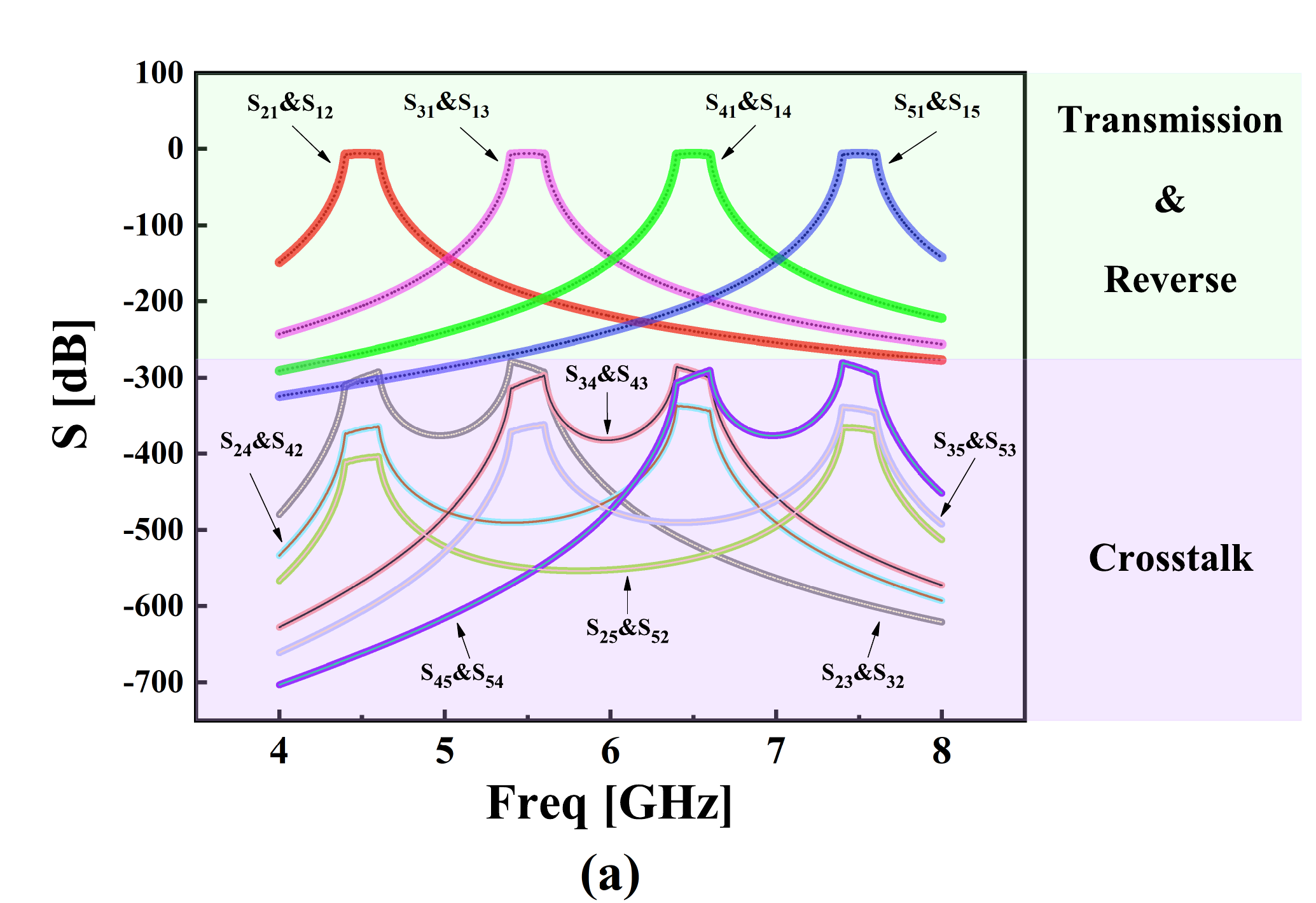}}
		\subfigure
		{\includegraphics[height=4.7cm,width=5.5cm]{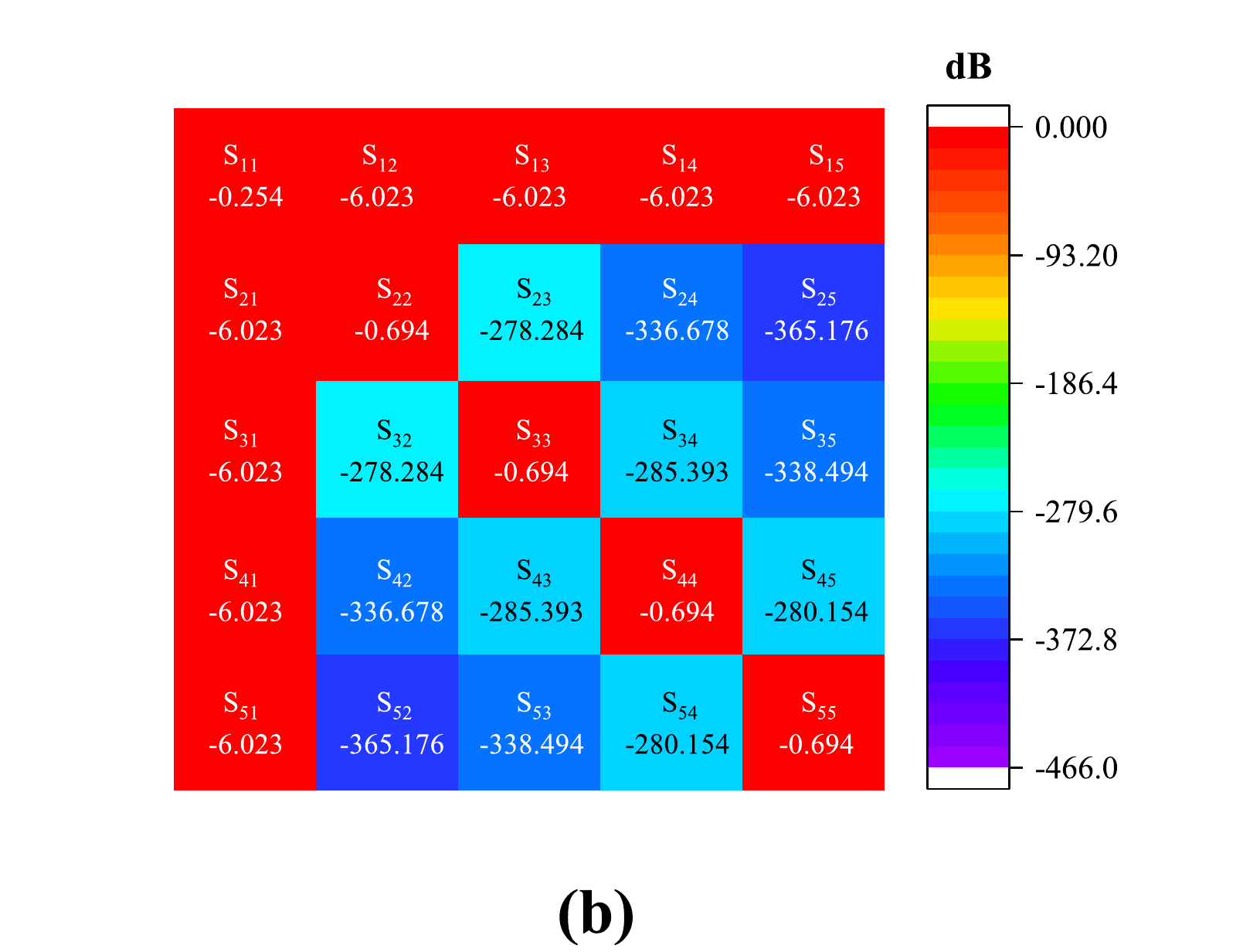}}
		\subfigure
		{\includegraphics[height=4.4cm,width=5cm]{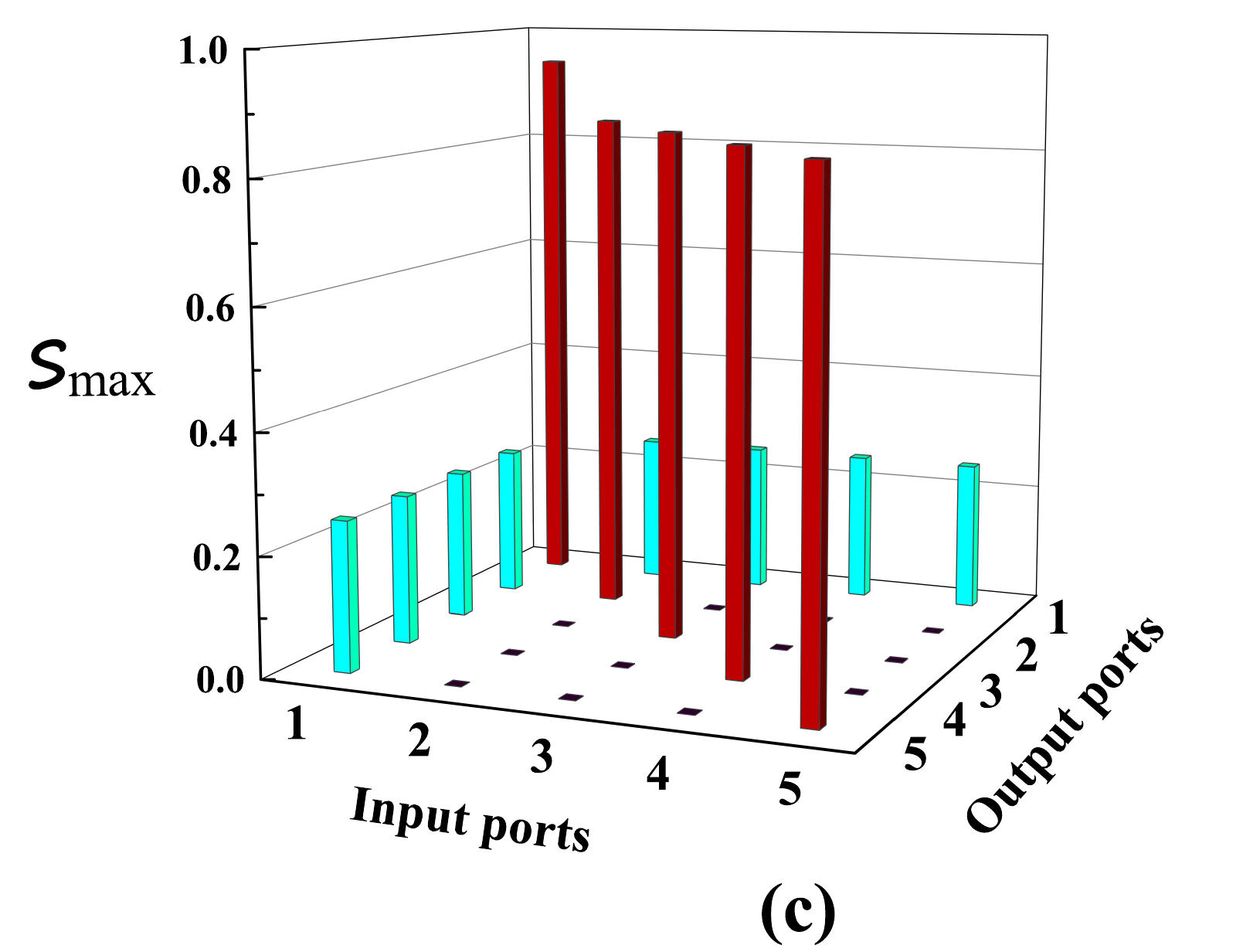}}
		{\includegraphics[height=4.6cm,width=7.2cm]{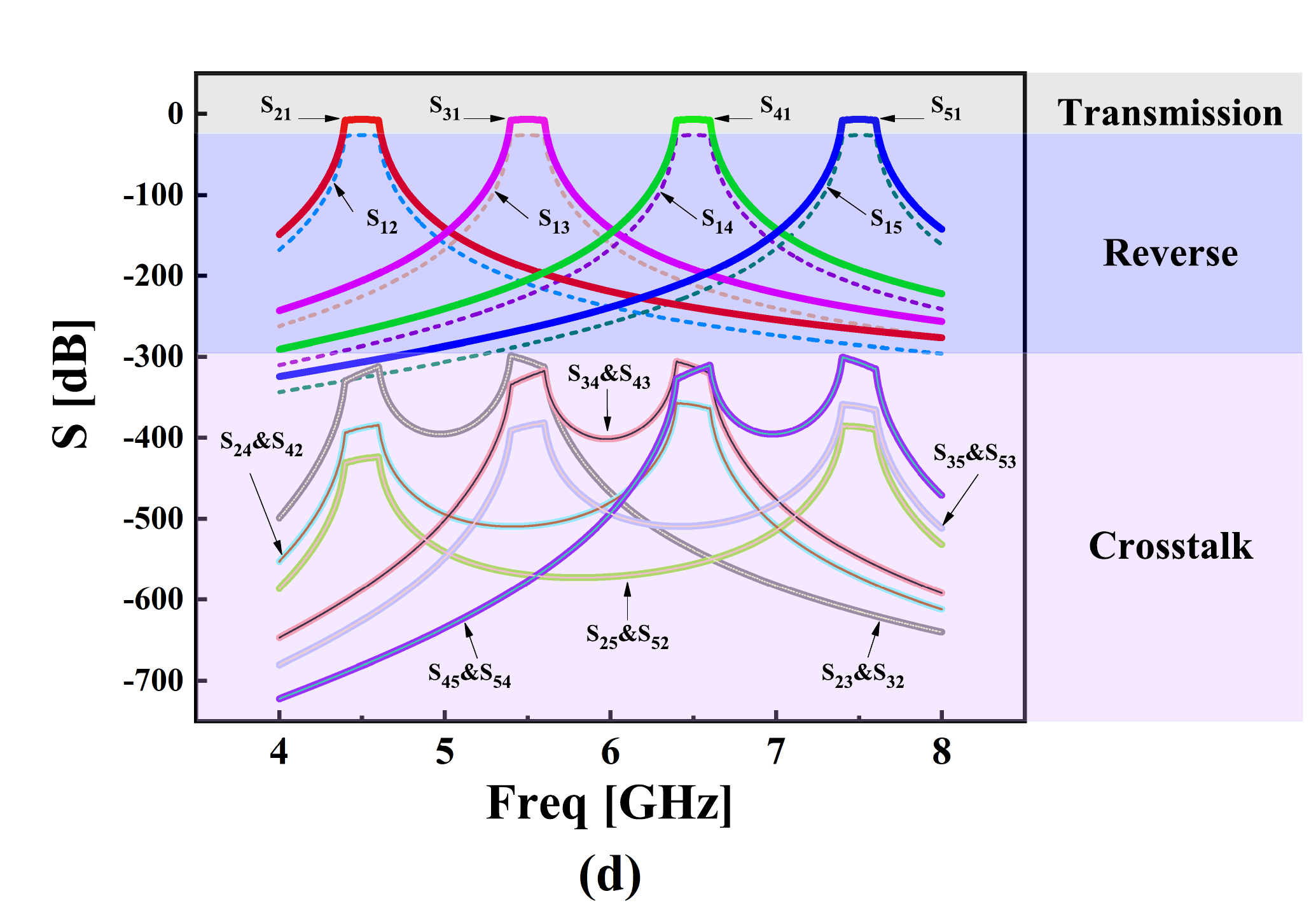}}
		\subfigure
		{\includegraphics[height=4.7cm,width=5.5cm]{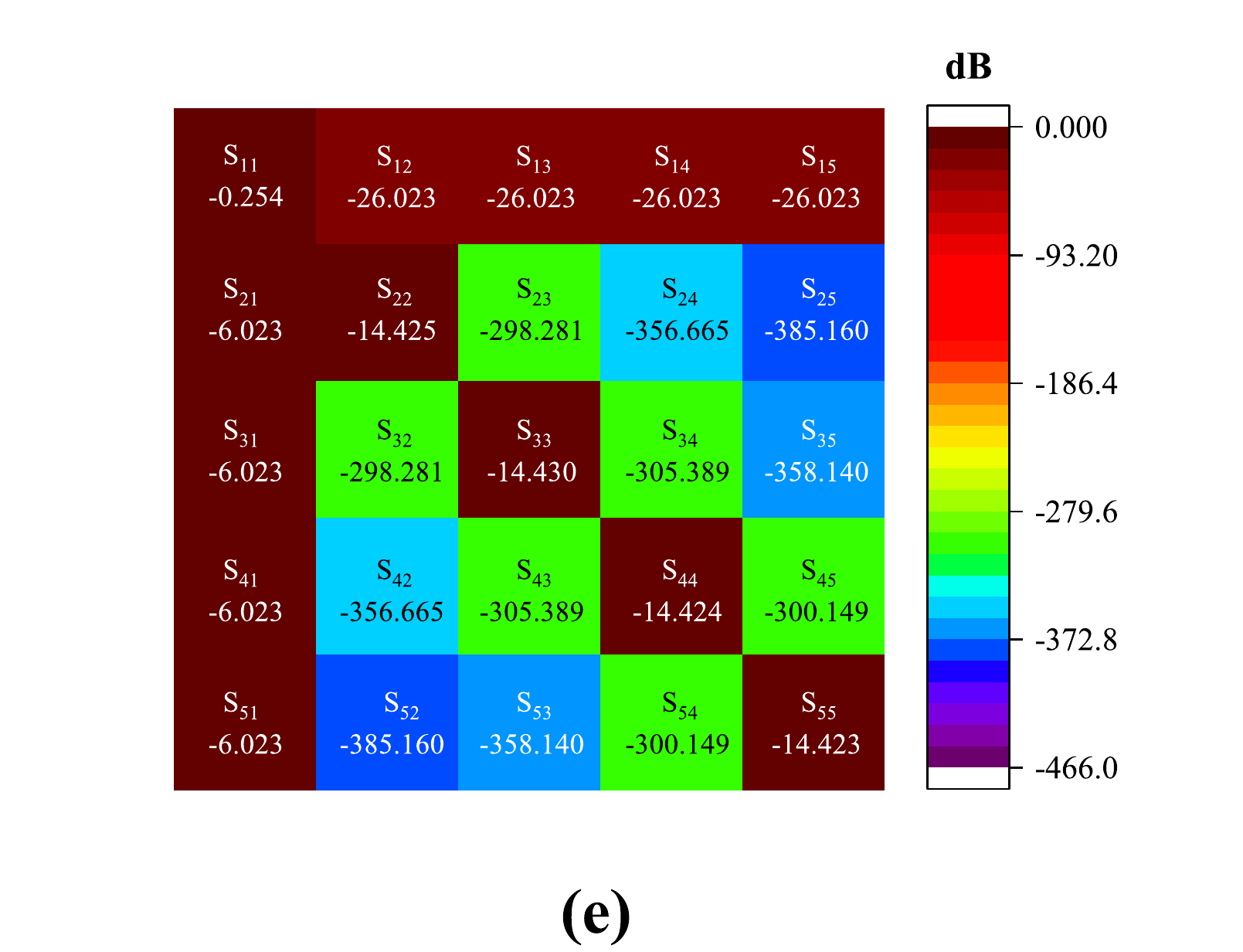}}
		\subfigure
		{\includegraphics[height=4.4cm,width=5cm]{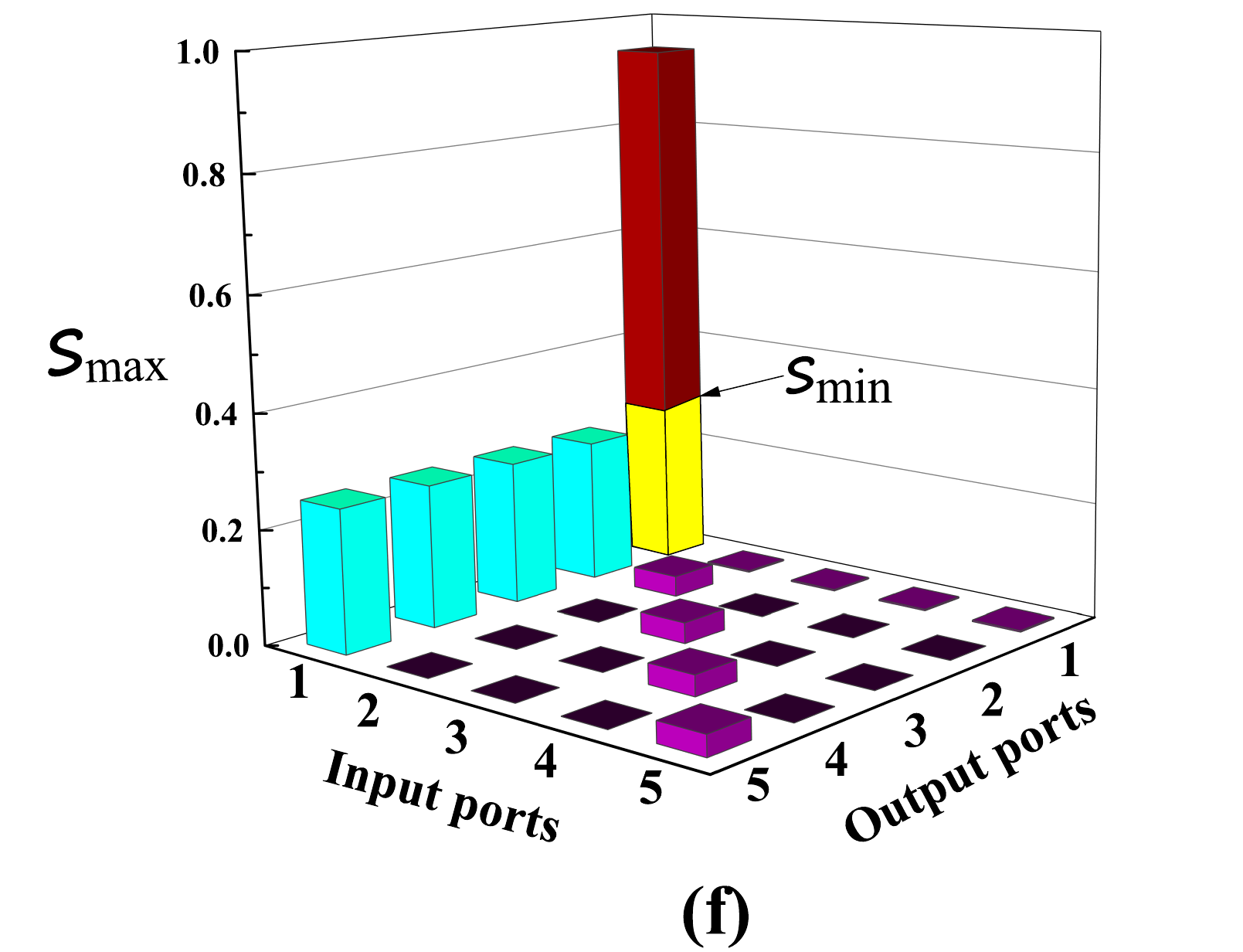}}
		\caption{The influence of crosstalks on MCC without isolator (a)(b)(c) and with isolators (d)(e)(f). \newline(a)(d) S parameter curves. $S_{\rm{max}}$ of the scattering matrix (b)(e) in dB, (c)(f) in linear unit.\linebreak The minimum of 
			$\mathcal{S}_{11}$ ($\mathcal {S}_{\rm{min}}$) is depicted as yellow pillar in (f).}
		\label{fig:F-crosstalk}
	\end{figure*}
	
	
	\subsection{Effects of qubits and impedance}
	\label{subsec:F-qubits-impedance}
	
	The stability of the MCC is critical for controling qubits. Superconducting qubits $Q_1$-$Q_4$, as shown on the right side of Fig.~\ref{fig:domain-simulation}, are designed with capacitors and   Josephson junctions which are modeled as nonlinear inductance. The inset of Fig.~\ref{fig:F-qubits-impedance} is a plot of $S_{\rm{max}}$ versus load impedance $Z_{\rm{load}}$, which can be fitted as
	\begin{equation}
	\begin{aligned}
	S_{\rm{max}} &=-0.274 Z_{\rm{load}}^4-1.799 Z_{\rm{load}}^3 \\
	&-0.264 Z_{\rm{load}}^2+11.047 Z_{\rm{load}}-17.664.
	\end{aligned}
	\end{equation}
	It is clear that the load impedance has great influence on S parameters, and the maximum point P(50 $\Omega$, $-$6 dB) indicates that $S_{\rm{max}}$ is stable at the matching load of 50 $\Omega$.
	
	The lower part in Fig.~\ref{fig:F-qubits-impedance} depicts the relationship between $S_{\rm{max}}$ and the inductance of qubits, $S_{\rm{max}}=-6$ dB, at the matching load impedance of 50 $\Omega$. When the inductance value in qubits is changed, it is found that the $S_{\rm{max}}$ does not change, which is a constant relation. Therefore, MCC devices are quite stable and would not be disturbed and affected by qubits parameters.
	
	\subsection{Reflection and crosstalk among channels}
	\label{subsec:F_crosstalk}
	
	The reflections and crosstalks among channels of MCC are characterised by simulating the correspongding S parameters. Fig.~\ref{fig:F-reflection} (a) gives the reflections of MCC without isolators. It can be seen that all of the reflections ($S_{11}$-$S_{55}$) are at the same level, and the minima appear at the center frequency of the filters. By putting the isolators on MCC, as shown in Fig.~\ref{fig:F-reflection} (b), the reflection $S_{11}$ remains the same, while others ($S_{22}$-$S_{55}$) are suppressed by about 14 dB.
	
	To clarify the effects of crosstalks, Fig.~\ref{fig:F-crosstalk} (a) presents the S parameters of MCC without isolators. The peaks of S parameters could be classified into two bands, i.e., transmission \& reverse band and crosstalk band. In the transmission \& reverse band, it is noticeable that $S_{i1}$=$S_{1i}$, $i$=2-5. In the crosstalk band, the S parameters ($S_{ij}$=$S_{ji}$: $i,j$=2-5, and $i \neq j$) are under $-270$ dB. Fig.~\ref{fig:F-crosstalk} (b) and (c) plot the $S_{\rm{max}}$ of the scattering matrix in dB and linear unit, respectively. It can be seen that they are symmetric matrix, which manifests the reciprocity of MCC without isolators.
	
	In the case of MCC with isolators, the S parameters are plotted in Fig.~\ref{fig:F-crosstalk} (d). The peaks of the S parameters could be separated approximately into three bands, i.e., transmission, reverse and crosstalk. The S parameters of the transmission ($S_{i1}$: $i$=2-5) is about $-6$ dB, while the reverse parameters ($S_{1j}$: $j$=2-5) are roughly below $-26$ dB. Furthermore, the S parameters of the crosstalks ($S_{ij}$: $i,j$=2-5, and $i \neq j$) are suppressed down even lower than $-300$ dB, which could be neglected safely. 
	
	Fig.~\ref{fig:F-crosstalk} (e) gives the $S_{\rm{max}}$ of the scattering matrix. The elements of the first red column ($S_{i1}$: $i$=1-5) are significantly larger than other elements by more than 8 dB. Thus,  the simulation results satisfy the design requriment of the S parameters as indicated by the ideal scattering matrix  Eq.\ref{ideal-s}, noticing that the unit of $S_{\rm{max}}$ is in dB here in Fig.~\ref{fig:F-crosstalk} (e). It can be seen that the S parameter matrix is asymmetric ($S_{i1} \neq S_{1i}$: $i$=2-5), so it can further verify the non-reciprocity of MCC devices. Fig.~\ref{fig:F-crosstalk} (f) presents a histogram of the $\mathcal{S}_{\rm{max}}$ matrix in the linear unit, and this is in agreement with the ideal scattering matrix Eq.\ref{ideal-s}.
	
	
	\section{TIME-DOMAIN SIMULATION}
	\label{sec:time_domain}
	
	
	\subsection{Response time}
	\label{subsec:Response time}
	
	In Sec. \ref{subsec:F_splitter}, the influence of power splitter on MCC has been compared and analyzed in the frequency domain simulation. Here, it is investigated further about response time with only the second channel (Ch.2) chosen. 
	
\begin{figure}[h]
		\begin{center}
			\includegraphics[height=6.425cm,width=8.6cm]{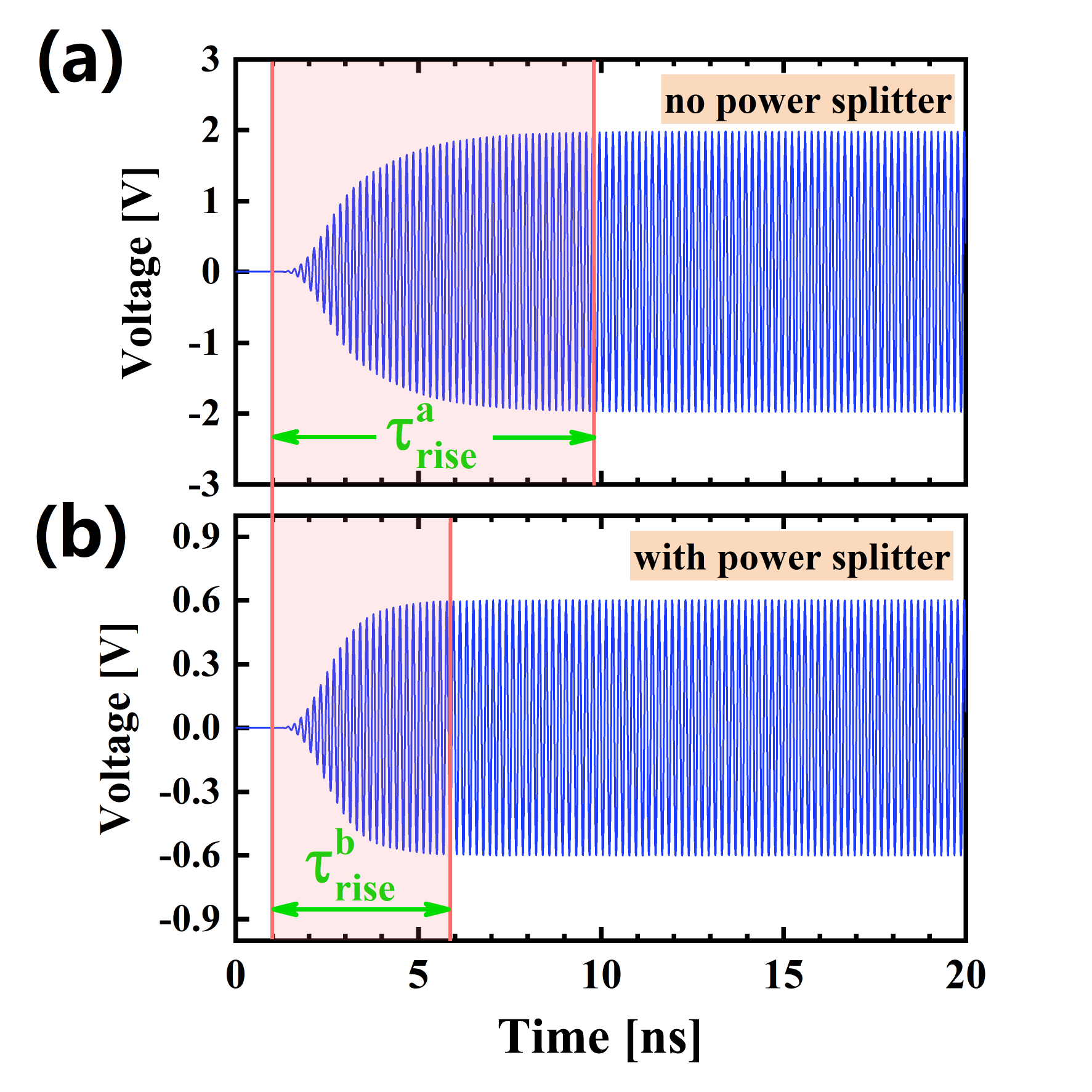}
			\caption{Effects of power splitter on response time of MCC. The simulations start at 1 ns for the continuous waves. The response time is determined from 1 ns to the time of the first maximum. (a) MCC without power splitters, $\tau_{\rm{rise}}^{\rm a}=9.961$ ns. (b) MCC with power plitters, $\tau_{\rm{rise}}^{\rm b}=5.959$ ns.}
			\label{fig:T-response-time-eps-converted-to}
		\end{center}
	\end{figure}

	Fig.~\ref{fig:T-response-time-eps-converted-to} (a) gives the continuous wave for the circuit without power splitters, and (b) for the circuit with power splitters. It can be seen that after adding the power splitter, the response time is shortened from $\tau_{\rm{rise}}^{\rm a}=9.961$ ns to $\tau_{\rm{rise}}^{\rm b}=5.959$ ns. This would be beneficial to carrying out faster quantum gates.
	
	\subsection{Continuous wave and Gaussian burst}
	\label{subsec:Gaussian}
	
	\begin{figure*}[t]
		\centering
		\subfigure
		{\includegraphics[height=8cm,width=8.5cm]{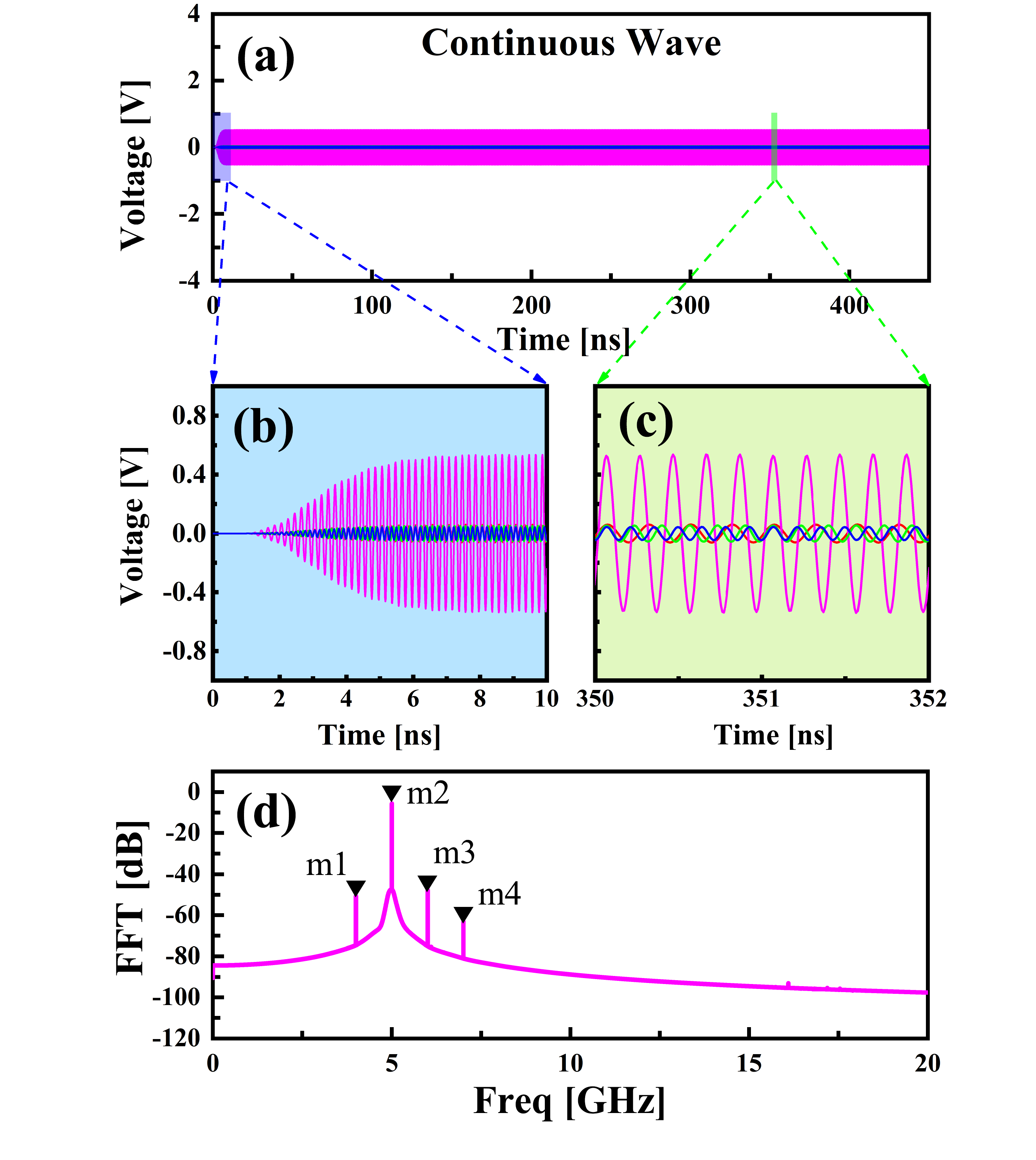}}
		\subfigure
		{\includegraphics[height=8cm,width=8.5cm]{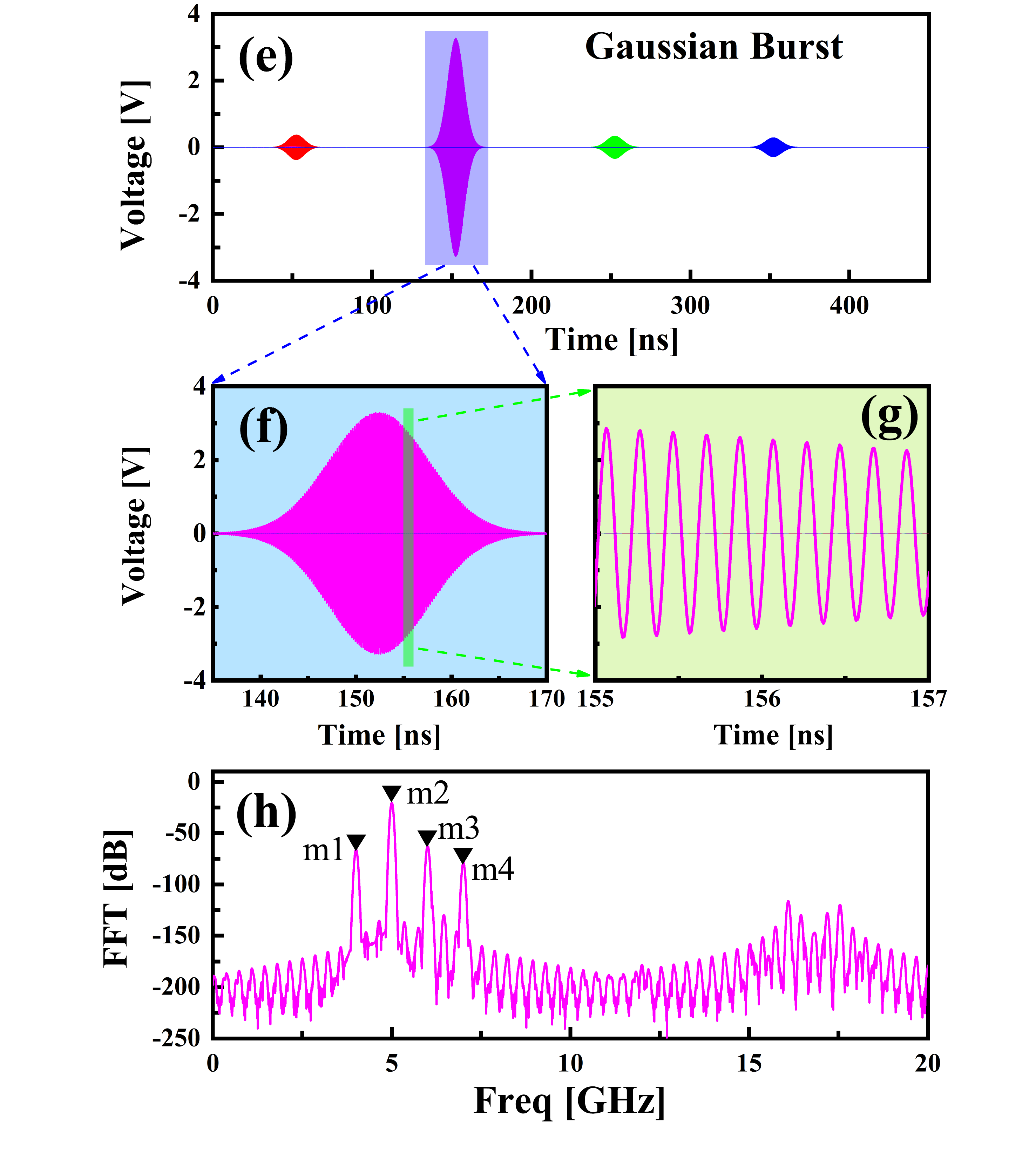}}
		\caption{Comparison of time domain simulations between continuous wave (left) and Gaussian shaped burst (right). The second channel on the MCC is made accessible, and the isolators of the other channels are reversed. 
			\newline(a) Continuous wave, (b) transient waveforms, (c) steady state waveforms, (d) FFT of the waveform in (a).\linebreak (e) Gaussian shaped burst, (f) Gaussian burst in the second channel, (g) zoom of waveform from 155 ns to 157 ns, (h) FFT of the waveform in (e). The details of peaks marked in (d) and (h) are listed in Table \ref{tab:Gaussian burst}.}
		\label{fig:T-continuous-Gaussian}
	\end{figure*}
	
	In superconducting quantum measurement systems, the microwave source usually generates continuous waves, and Gaussian shaped bursts are explored to control qubits. As mentioned in Sec. \ref{subsec:Time-domain model analysis} for IQ mixer, the functions of Gaussian in $I(t)$ and $Q(t)$ are
	\begin{equation}
	I(t)=Q(t)=ae^{-\frac{(t-t_i)^2}{2 \sigma^2}},
	\label{f_I}
	\end{equation}
	where $i=1$-4, 
	$a=10$ \rm{V},
	$t_1=50$ ns,
	$t_2=150$ ns,
	$t_3=250$ ns,
	$t_4=350$ ns,
	$\sigma=5$ ns,
	$t$ is from 0 to 450 ns with step size of 1 ns.
	
	Fig.~\ref{fig:T-continuous-Gaussian} shows the comparison of simulation results for continuous wave and Gaussian shaped burts. Here, the source frequencies of four channels are 4.000, 5.000, 6.000 and 7.000 GHz, respectively. The second channel on the MCC is made accessible, and the isolators of the other channels are reversed to block it. Fig.~\ref{fig:T-continuous-Gaussian} (a) displays the continuous wave, while (b) and (c) are transient waveform and the steady state waveforms, respectively. Fig.~\ref{fig:T-continuous-Gaussian} (d) is the frequency spectrum obtained from the fast Fourier transformation (FFT) of the waveform  in Fig.~\ref{fig:T-continuous-Gaussian} (a). Similarly, Fig.~\ref{fig:T-continuous-Gaussian} (e) displays the Gaussian shaped bursts, while (f) is the Gaussian burst for the second channel and (g) is a stretch of 2 ns of the waveform between 155 ns and 157 ns, respectively. Fig.~\ref{fig:T-continuous-Gaussian} (h) is the FFT of the waveform  in Fig.~\ref{fig:T-continuous-Gaussian} (e). The peaks in Fig.~\ref{fig:T-continuous-Gaussian} (d) and (h) are listed in Tab \ref{tab:Gaussian burst}. It can be seen that the frequency of each channel is precise for continuous wave, and the frequency of each channel is slightly deviated ($\sim$ $\pm4$ MHz) from the source frequency when using the Gaussian burst. Such a deviation of frequency is acceptable for controling qubits.
	
	\begin{table}[h]
		\renewcommand\arraystretch{2}
		\centering
		\caption{Peaks of FFT for continuous wave and Gaussian burst.}
		\label{tab:Gaussian burst}
		\begin{tabular}{C{0.8cm}C{1.8cm}C{2.5cm}C{2.5cm}}
			\hline
			\hline
			\multirow{2}{*}{$\rm{Peak}$} & Source  &   Continuous wave & Gaussian burst  \\ 
			&(GHz)   &(GHz, dB)       &(GHz, dB)\\
			\hline
			$\rm{m_1}$  & 4.000 & (4.000, $-52.074$)     & (4.004, $-67.928$) \\
			$\rm{m_2}$  & 5.000 & (5.000, $-5.595$)      & (4.998, $-20.627$) \\
			$\rm{m_3}$  & 6.000 & (6.000, $-49.500$)     & (5.998, $-64.621$) \\
			$\rm{m_4}$  & 7.000 & (7.000, $-64.828$)     & (6.998, $-80.450$) \\ 
			\hline
			\hline
		\end{tabular}
	\end{table}
	
	
	\subsection{Single-sideband}
	\label{subsec:T_SSB}
	
	\begin{figure*}[t]
		\centering
		\subfigure
		{\includegraphics[height=12cm,width=5.6cm]{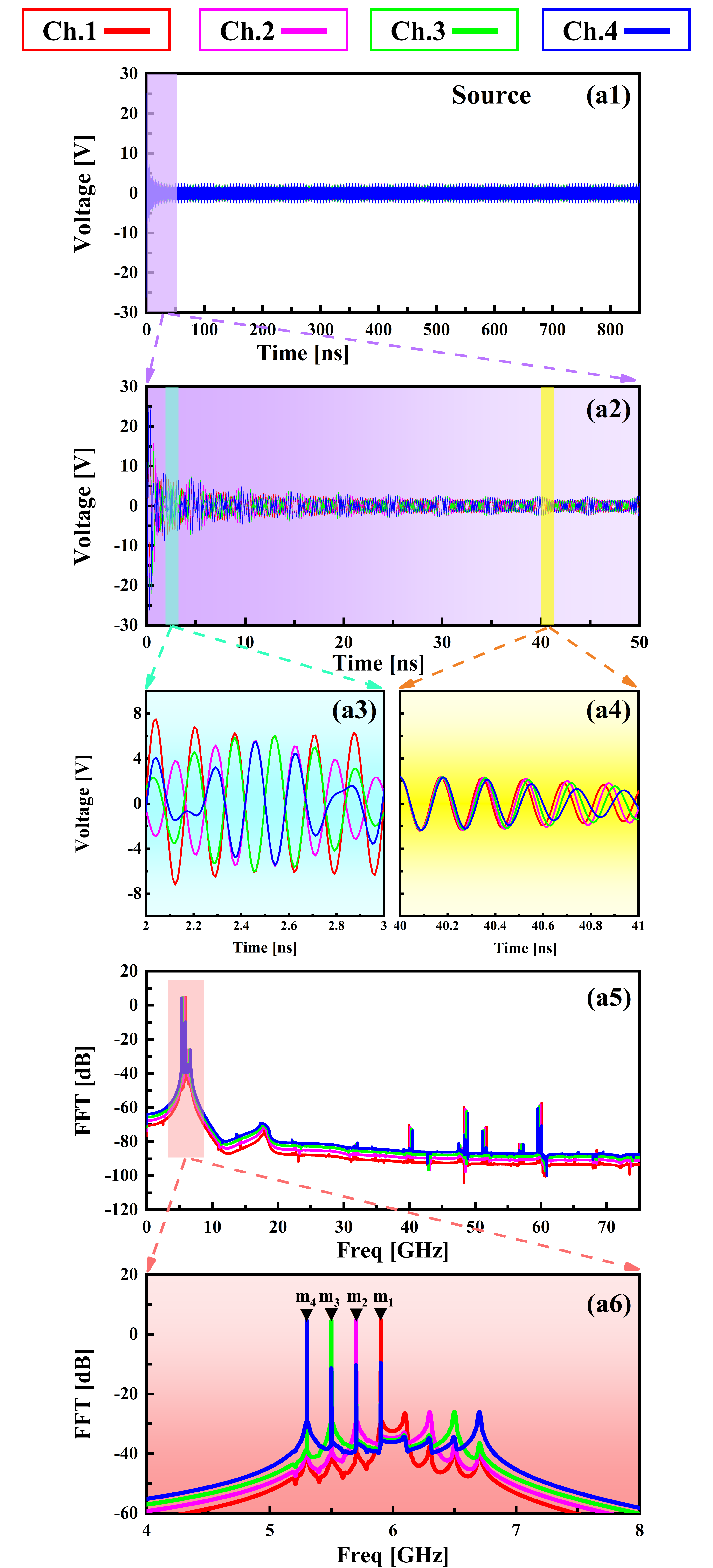}}
		\subfigure
		{\includegraphics[height=12cm,width=5.6cm]{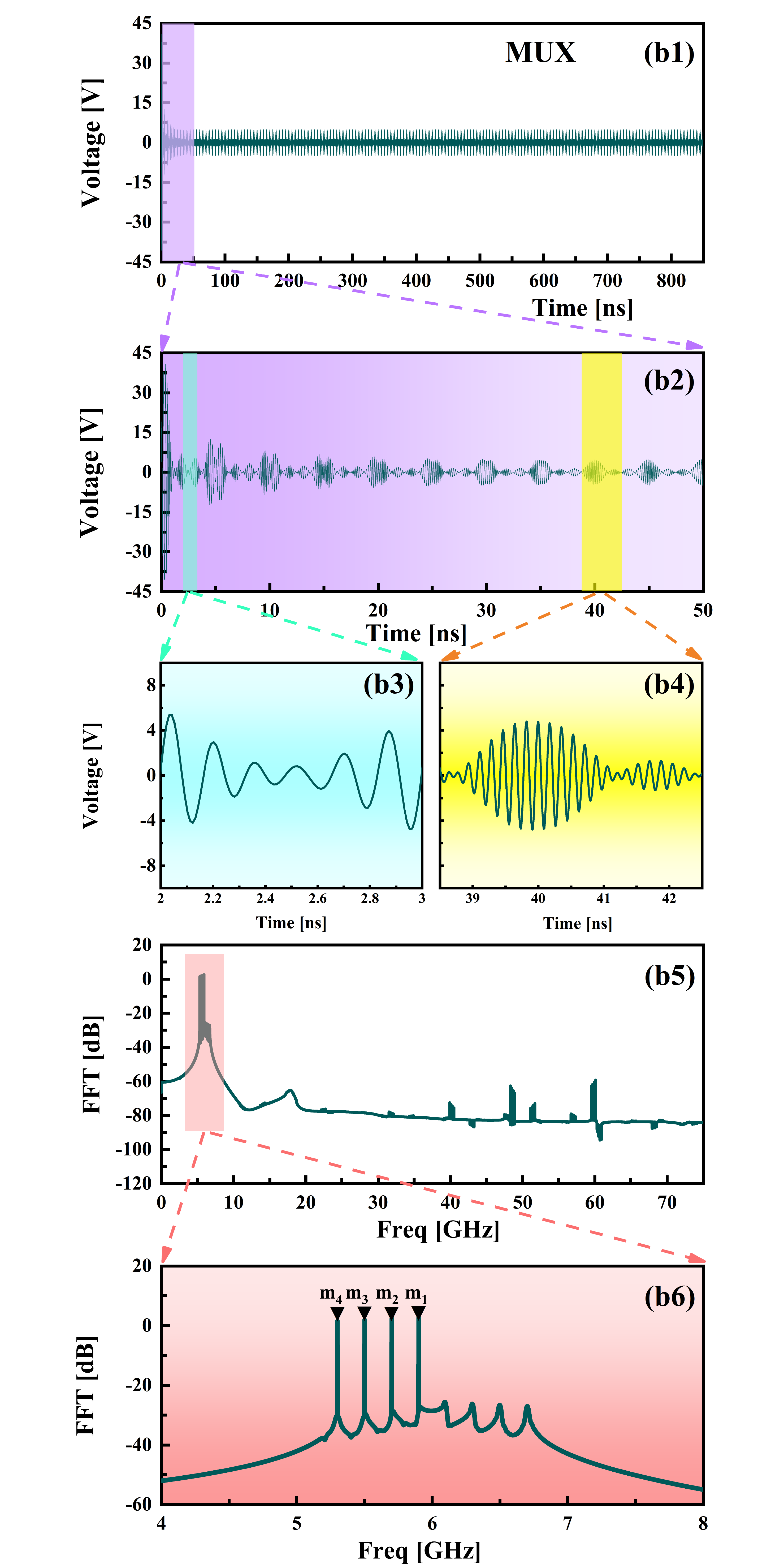}}
		\subfigure
		{\includegraphics[height=12cm,width=5.6cm]{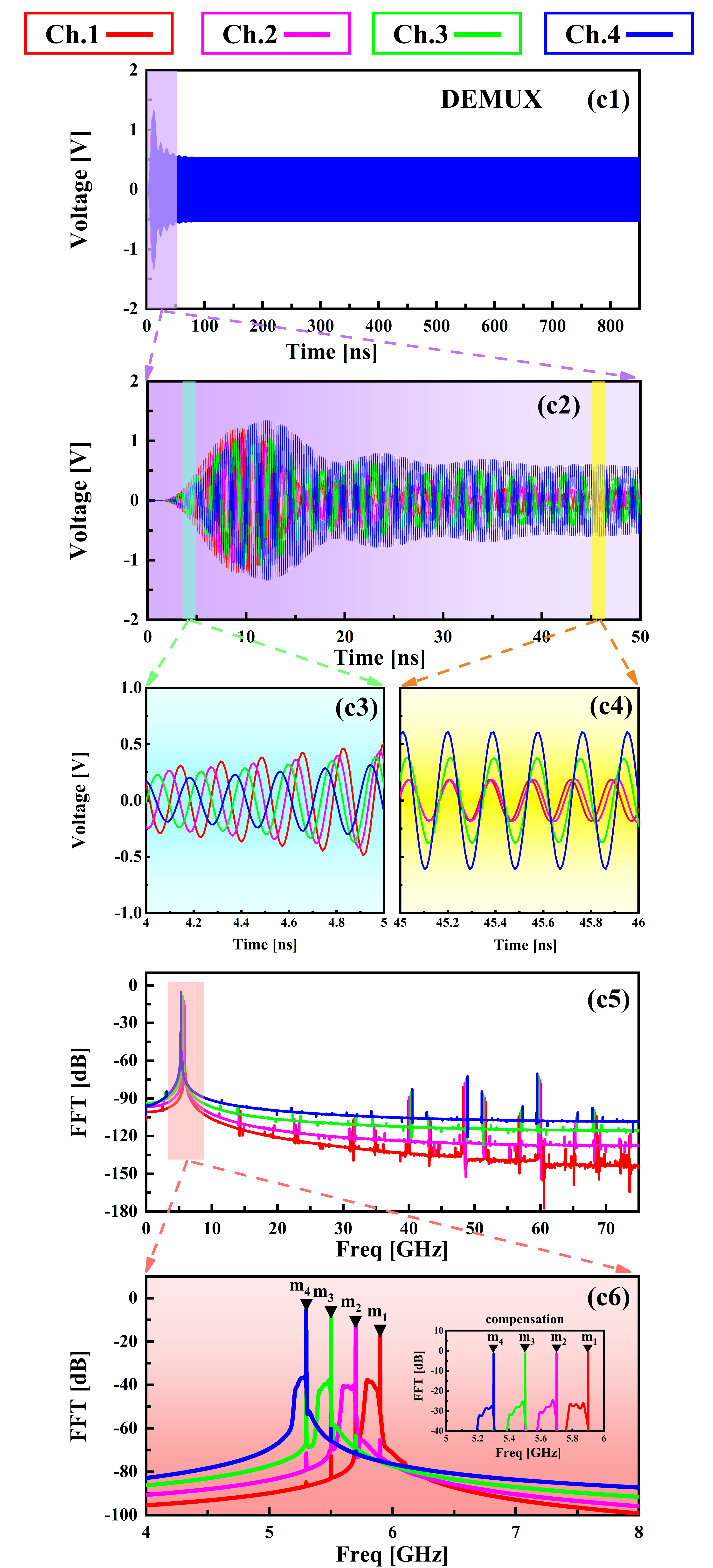}}
		\caption{The simulation results of SSB with continuous waves for source (left), multiplex (middle) and demultiplex (right). Here all four channels are accessible. (a1) Source; (a2) transient waveforms; (a3) and (a4) are segments in (a2); (a5) FFT of the source; (a6) detailed contents in the range of 4-8 GHz. (b1) MUX; (b2) transient waveforms; (b3) and (b4) are segments in (b2); (b5) FFT of the MUX; (b6) detailed contents highlighted in (b5). (c1) DEMUX; (c2) transient waveforms; (c3) and (c4) are segments in (c2); (c5) FFT of the DEMUX; (c6) detailed contents highlighted in (c5); inset of (c6) is the spectrum with amplitude compensation. The details of peaks marked in (a6), (b6) and (c6) are listed in Table \ref{tab:No Gaussian single-sideband}.}
		\label{fig:T-NG-SSB}
	\end{figure*}
	
	In time domain analysis of MCC, single-sideband (SSB) is also adopted. As mentioned in Sec. \ref{subsec:Time-domain model analysis} for IQ mixer, the $I(t)$ and $Q(t)$ are 
	\begin{equation}
	I(t)= \cos(2 \pi f_i t), 
	\end{equation}
	\begin{equation}
	Q(t)= \sin(2 \pi f_i t),
	\end{equation}
	where $i=1$-4, 
	$f_1=100$ MHz,
	$f_2=300$ MHz,
	$f_3=500$ MHz,
	$f_4=700$ MHz,
	$t$ is from 0 to 850 ns, step size is 10 ns.
	
	A microwave source as LO signal with one frequency passes through power splitters and IQ mixers via SSB to generate multiple microwave sources with different frequencies, which greatly saves the number of microwave source. In this way, phase coherence of qubit control is attainable \cite{Arute2019}. 
	
	In this work, four channels of microwave source with different frequencies  are combined in one microwave line through power splitters. That is MUX. Then, the microwave signals are demultiplexed and accurately separated into four independent channels through MCC. This process is called DEMUX. System-level simulations of MUX and DEMUX are conducted for continuous wave and Gaussian burst, and FFT is performed on the time domain results for the frequency spectrum analysis.
	
	Fig.~\ref{fig:T-NG-SSB} displays the simulation results of SSB with continuous waves, where all four channels are accessible. Here, (a1-6) are the waveforms and FFT for the source, (b1-6) are the waveforms and FFT for the multiplexed curves, and (c1-6) are those for the demultiplexing. It is noticeable that, in Fig.~\ref{fig:T-NG-SSB} (a3) and (a4), there are four traces corresponding to the four channels in the source. In Fig.~\ref{fig:T-NG-SSB} (b3) and (b4), these four traces are converged into a single one through MUX, which is separated back to four traces via DEMUX in Fig.~\ref{fig:T-NG-SSB} (c3) and (c4). The frequencies are conserved precisely during the signal processing based on Fig.~\ref{fig:T-NG-SSB} (a6, b6, c6) and Table \ref{tab:No Gaussian single-sideband}. In addition, although amplitudes of the four channels are the same in (a6) and (b6), they are different from each other in (c6). These variations on the amplitudes could be compensated accordingly by adjusting the amplitude of the corresponding sources.
	
	\begin{table*}[h]
		\renewcommand\arraystretch{2}
		\centering
		\caption{Peaks of FFT for the case of continuous wave with SSB.}
		\label{tab:No Gaussian single-sideband}
		\begin{tabular}{C{2cm}C{3.5cm}C{3.5cm}C{3.5cm}C{3.5cm}}
			\hline
			\hline
			\multirow{2}{*}{$\rm{Peak}$}  & Source   & MUX    & DEMUX   & Compensation    \\ 
			&(GHz, dB) &(GHz, dB) &(GHz, dB)  &(GHz, dB)\\
			\hline
			$\rm{m_1}$   & (5.900, 4.896) & (5.900, 2.724) & (5.900, $-16.665$) & (5.900, $-1.252$) \\
			$\rm{m_2}$    & (5.700, 4.779) & (5.700, 2.380) & (5.700, $-12.799$) & (5.700, $-1.295$) \\
			$\rm{m_3}$    & (5.500, 4.638) & (5.500, 1.984) & (5.500, $-8.872$)  & (5.500, $-1.229$) \\
			$\rm{m_4}$    & (5.300, 4.514) & (5.300, 1.614) & (5.300, $-5.463$)  & (5.300, $-1.220$) \\ 
			\hline
			\hline
		\end{tabular}
	\end{table*}
	
	\begin{figure*}[h]
		\centering
		\subfigure
		{\includegraphics[height=12cm,width=5.5cm]{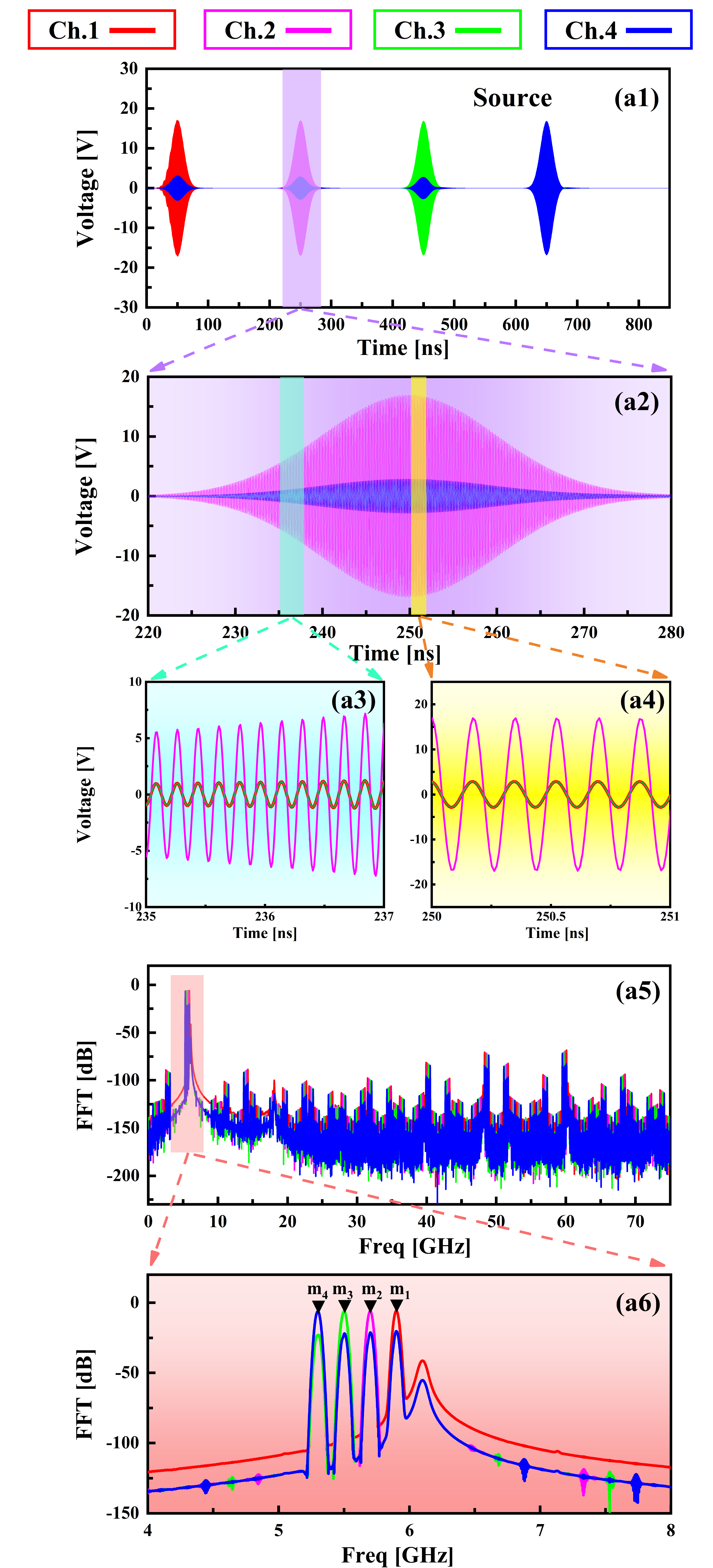}}
		\subfigure
		{\includegraphics[height=12cm,width=5.5cm]{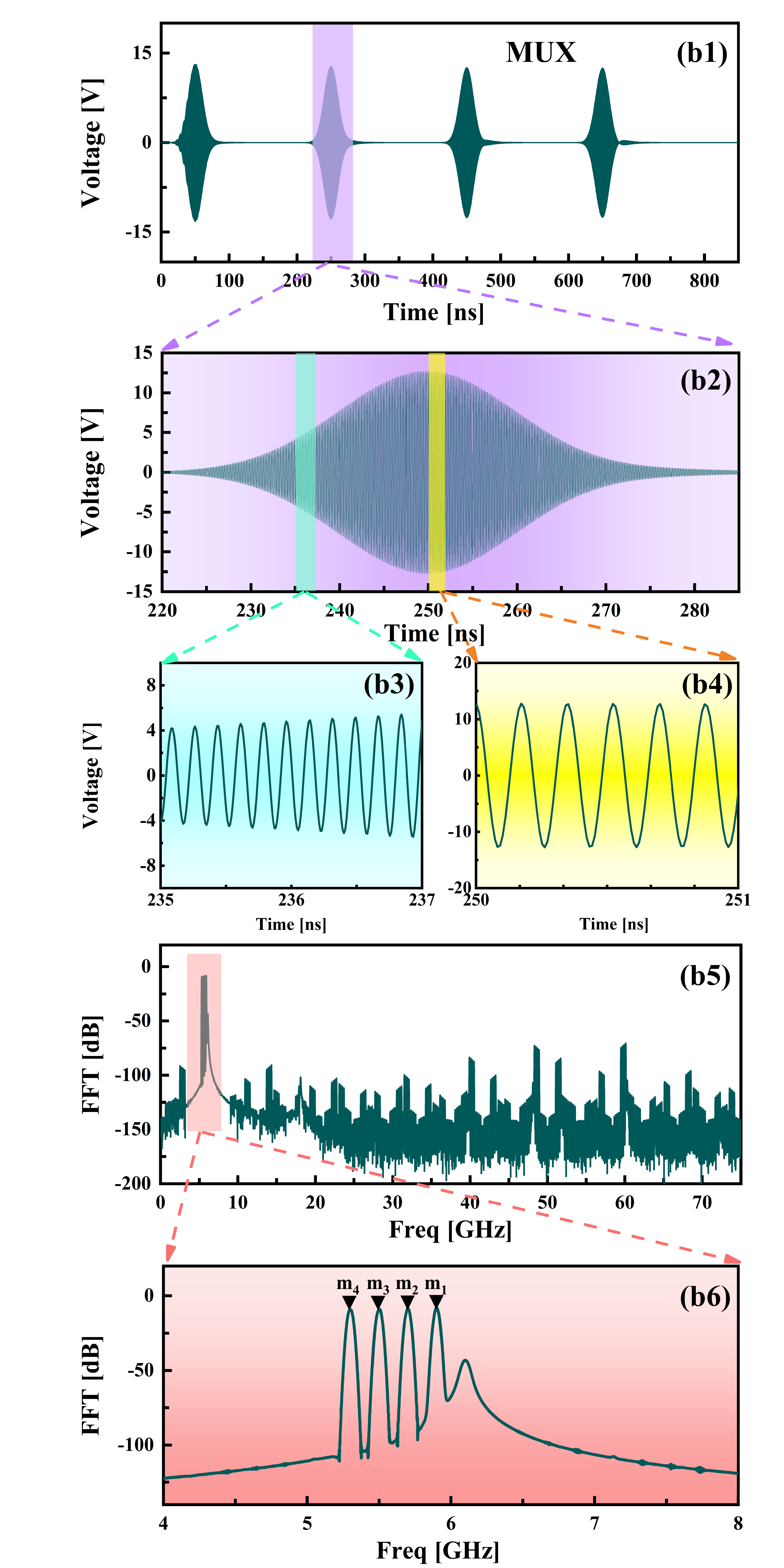}}
		\subfigure
		{\includegraphics[height=12cm,width=5.5cm]{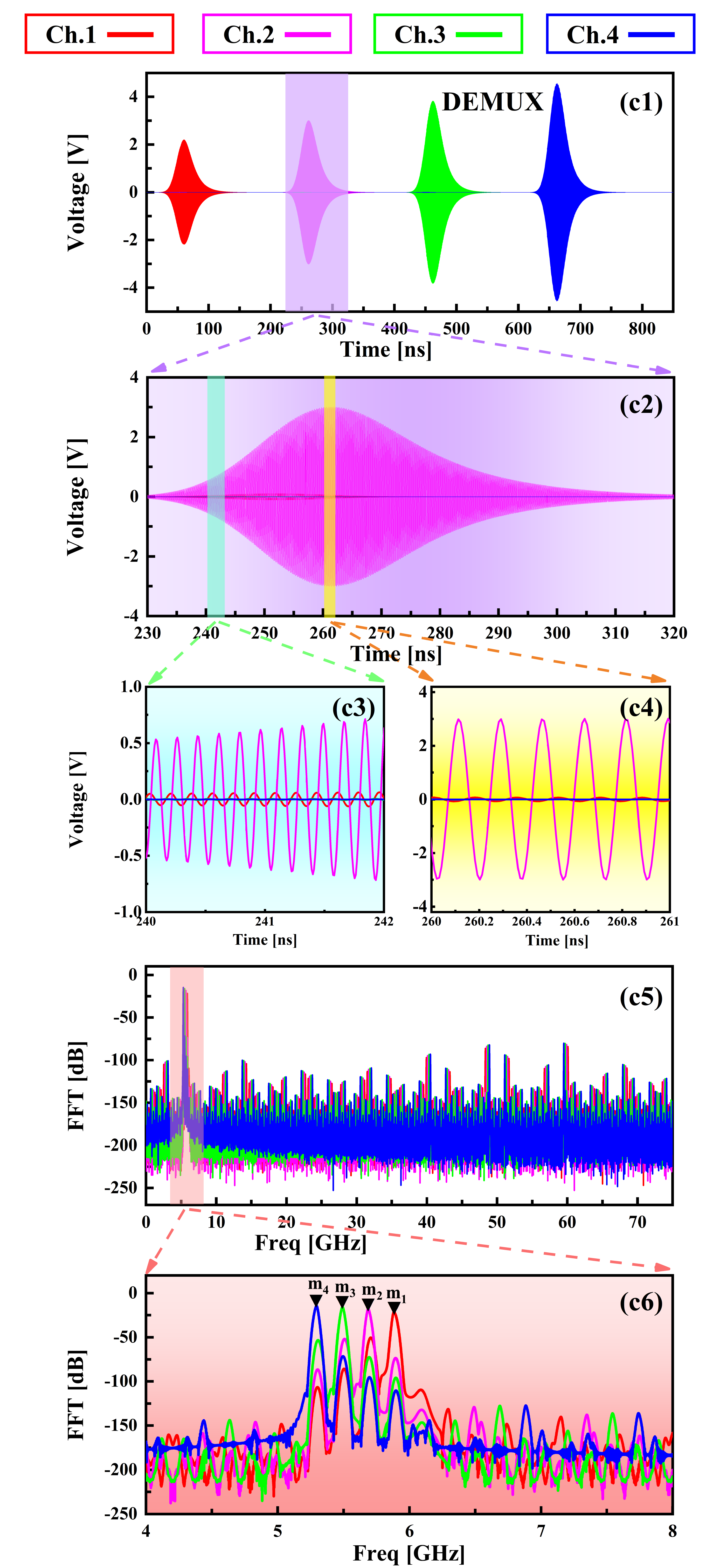}}
		\caption{The simulation results of SSB with Gaussian bursts for source (left), multiplex (middle) and demultiplex (right). Here all four channels are accessible, but attention is focused on the second channel. (a1) Source; (a2) the Gaussian pulse in Ch.2; (a3) and (a4) are enlarged parts of (a2); (a5) FFT of the source; (a6) detailed contents in the range of 4-8 GHz. (b1) MUX; (b2) the second Gaussian pulse; (b3) and (b4) are enlarged parts of (b2); (b5) FFT of the MUX; (b6) detailed contents highlighted in (b5). (c1) DEMUX; (c2) the pulse in Ch.2; (c3) and (c4) are enlarged parts of (c2); (c5) FFT of the DEMUX; (c6) detailed contents highlighted in (c5). The details of peaks marked in (a6), (b6) and (c6) are listed in Table \ref{tab:Gaussian single-sideband}.}
		\label{fig:T-WG-SSB}
	\end{figure*}
	
	To have Gaussian shaped pulses, $I(t)$ and $Q(t)$ of IQ mixers are applied respectively, that is:
	\begin{equation}
	I(t)=ae^{-\frac{(t-t_i)^2}{2 \sigma^2}} \cos(2\pi f_i t),
	\label{f_I}
	\end{equation}
	\begin{equation}
	Q(t)=ae^{-\frac{(t-t_i)^2}{2 \sigma^2}}  \sin(2\pi f_i t),
	\label{f_Q}
	\end{equation}
	where $i=1$-4, 
	$a=10$ \rm{V},
	$t_1=50$ ns,
	$t_2=250$ ns,
	$t_3=450$ ns,
	$t_4=650$ ns,
	$\sigma=10$ ns,
	$f_1=100$ MHz,
	$f_2=300$ MHz,
	$f_3=500$ MHz,
	$f_4=700$ MHz,
	$t$ is from 0 to 850 ns, step size is 10 ps.  
	
	Fig.~\ref{fig:T-WG-SSB} illustrates the simulation results of SSB with Gaussian bursts for source (a1-6), multiplex (b1-6) and demultiplex (c1-6). Here all four channels are accessible, but attention is focused on the second channel. It could be seen that, in Fig.~\ref{fig:T-WG-SSB} (a1), Gaussian pulses in the four channels are arranged with a time interval of  about 200 ns, which is also maintained in (b1) and (c1). The voltage amplitudes of four channels in (a1) are the same, and similiarly in (b1),  while differing from each other in (c1). Furthermore, the dropping of voltage peak in the Source-MUX-DEMUX chain is caused by energy dissipation on the power splitters and filters. In the superconducting quantum test system, the attenuation of controling signals is necessary before the qubit operation. Thus some of the attenuators could be omitted accordingly to simplify the measurement system. 
	
	\begin{table}[h]
		\renewcommand\arraystretch{1.6}
		\centering
		\caption{Peaks of FFT for the case of Gaussian bursts with SSB.}
		\label{tab:Gaussian single-sideband}
		\begin{tabular}{C{0.8cm}C{2.3cm}C{2.3cm}C{2.5cm}}
			\hline
			\hline
			\multirow{2}{*}{$\rm{Peak}$} & Source  & MUX     & DEMUX \\
			&(GHz, dB) &(GHz, dB)  &(GHz, dB)\\
			\hline
			$\rm{m_1}$   & (5.900, $-5.839$) & (5.900, $-8.023$) & (5.884, $-21.506$) \\
			$\rm{m_2}$   & (5.700, $-5.996$) & (5.700, $-8.426$) & (5.687, $-18.233$) \\
			$\rm{m_3}$   & (5.500, $-6.142$) & (5.500, $-8.839$) & (5.492, $-16.014$) \\
			$\rm{m_4}$   & (5.300, $-6.272$) & (5.302, $-9.182$) & (5.295, $-14.685$) \\ 
			\hline
			\hline
		\end{tabular}
	\end{table}
	
	It is noticeable that the Gaussian shapes are symmetric in Fig.~\ref{fig:T-WG-SSB} (a2) and (b2), which is broken in (c2). In Fig.~\ref{fig:T-WG-SSB} (a3, a4), there are four traces corresponding to the four channels in the Source. However, Ch.2 is higher than the other three coincident traces of Ch.1, Ch.3 and Ch.4. These four traces are united into a single one in (b3) and (b4), which is devided back again to four traces in Fig.~\ref{fig:T-WG-SSB} (c3) and (c4). Fig.~\ref{fig:T-WG-SSB} (a5, b5, c5) give the FFT of the waveforms in (a1, b1, c1), respectively, and (a6, b6, c6) are the zoom of detailed frequency spectra within 4-8 GHz. The peaks in Fig.~\ref{fig:T-WG-SSB} (a6, b6, c6) are listed in Table \ref{tab:Gaussian single-sideband}. In contrast to the case with continuous waves as shown in Fig.~\ref{fig:T-NG-SSB} (a6, b6, c6) and Table \ref{tab:No Gaussian single-sideband}, where the frequencies remain the same during the signal processing stages, the frequencies through DEMUX in Table \ref{tab:Gaussian single-sideband} have a redshift of about 5-16 MHz relative to the source frequencies. These frequency redshifts could also be compensated accordingly by manipulating the frequencies of the corresponding sources if needed \cite{Rol2020}.

	
	\section{CONCLUSION}
	\label{sec:conclusion}
	
	In summary, this paper presents a design of the cryogenic multiplexing control chip based on adjustable inductor to control superconducting qubits. MCC is a non-reciprocal device consisted of filters and isolators which demultiplexes microwave signals between 4 and 8 GHz at 20 mK temperature stage. The feasibility of the MCC device is verified by frequency domain analysis and time domain simulation. The effects of filters, isolators, power splitters, qubits and load impedance on MCC are examined in the frequency-domain analysis. The results on the crosstalks of S parameters have revealed the non-reciprocity of MCC which agrees with the ideal scattering matrix. The time-domain simulation has demonstrated single-sideband generation, multiplex and demultiplex of continuous waves and Gaussian shaped bursts.  Enclosed with the superconducting qubits in the same package or even on the same die, the MCC could distribute specific microwave pulses to drive the superconducting quantum processor without being interfered by crosstalks among various qubit channels. Combining with cryo-CMOS and other controlling techniques, this approach would enable a handful of microwave cables to control millions of superconducting qubits. 
	
	
	\section*{Acknowledgments}
	We acknowledge Changhao Zhao, Kaiyong He, Yunfan Shi, Qing Yu, and Yongcheng He for stimulating and valuable discussions. This work is partially supported by the National Natural Science Foundation of China (Grant Nos. 11653001, 11653004, and 60836001), and key R\&D program of Guangdong province (Grant No. 2019B010143002).
	
	\renewcommand
	\refname
	\bibliographystyle{}

\end{document}